\documentclass[preprint,prfluids,eqsecnum]{revtex4-2} % reprint 
\usepackage{natbib}
\usepackage{graphicx}
\usepackage{subfigure}
\usepackage{amssymb}
\usepackage{amsfonts}
\usepackage{amsmath}
\usepackage{amsbsy}
\usepackage{mathrsfs} 
\usepackage{color}
\usepackage[colorlinks=true,citecolor=blue]{hyperref} 
\usepackage{soul}%\st{...}
\usepackage{lineno}
\usepackage{graphicx}

\usepackage[normalem]{ulem} % Strikeout \sout
\newcommand{\os}[2][]{\textcolor{blue}{\ifx\relax#1\relax\else\sout{#1}\fi #2}} 
\newcommand{\osR}[2][]{\textcolor{red}{\ifx\relax#1\relax\else\sout{#1}\fi #2}} 
\newcommand{\intd}[1]{d#1} % Integral d
\newcommand{\pd}[2]{\frac{\partial #1}{\partial #2}}
\newcommand{\ee}{e}
\newcommand{\ii}{i}
\newcommand{\bu}{\mathbf{u}}
\newcommand{\bcdot}{\boldsymbol{\cdot}}
\newcommand{\be}{\hat{\mathbf{e}}}
\DeclareMathOperator{\ord}{ord}

\begin{document} 
%\linenumbers
\begin{abstract}
We consider shear-driven longitudinal flow of an exterior fluid over a periodic array of rectangular grooves filled with an immiscible interior fluid (the ``lubricant''), the grooves being formed by infinitely thin ridges protruding from a flat substrate. The ratio $\lambda$ of the effective slip length to the semi-period is a function of the ratio $\mu$ of the interior to exterior viscosities and the ratio $h$ of the grooves depth to the semi-period. We focus on the limit $\mu\ll1$, which is singular for that geometry. We find that the viscous resistance to the imposed shear is dominated by a boundary layer of \emph{exponentially small} extent about the ridge tips, resulting in the effective slip length scaling as $\mu^{-1/2}$ --- not $\mu^{-1}$ as implied by intuitive arguments overlooking the tip contributions (and by proposed approximations in the literature). Analyzing that exponential region in conjunction with an integral force balance, we find the simple asymptotic approximation $\lambda\approx \mu^{-1/2}$; using conformal mappings, we also calculate the leading-order correction to that result, which introduces a dependence upon $h$. The ensuing asymptotic expansion breaks down for $h=O(\mu^{1/2})$, upon transitioning to a lubrication geometry. We accordingly conduct a companion asymptotic analysis in the distinguished limit of small $\mu$ and fixed $H=h/\mu^{1/2}$, which gives $\lambda\approx \mu^{-1/2}H/(1+H)$ as well as a closed-form leading-order correction to that approximation; the intuitive $\mu^{-1}$ scaling is accordingly only relevant to the regime $H\ll1$ corresponding to extremely shallow grooves. We demonstrate excellent agreement between our predictions and numerical solutions constructed using a boundary-integral formulation.
\end{abstract}

\title{Slip over liquid-infused gratings in the \texorpdfstring{\\}{} singular limit of a nearly inviscid lubricant}
\author{Gunnar G. Peng$^1$, Ehud Yariv$^2$, Ory Schnitzer$^3$}
\affiliation{$^1\!$Department of Mathematics, University College London, London WC1E 6BT, United Kingdom}
\affiliation{$^2\!$Department of Mathematics, Technion---Israel Institute of Technology, Haifa 32000, Israel}
\affiliation{$^2\!$Department of Mechanical and Aerospace Engineering, Princeton University, Princeton, New Jersey 08544, USA}
\affiliation{$^3\!$Department of Mathematics, Imperial College London, London SW7 2AZ, United Kingdom}

%\date{\today}

\maketitle
\newpage
\section{Introduction}
There is continued interest in the use of engineered microstructured surfaces to reduce hydrodynamic resistance at small scales. Such drag reduction typically hinges upon the stable entrapment of a relatively inviscid lubricant within the microstructure  cavities. Traditionally, research has focused on superhydrophobic surfaces \citep{Rothstein:10,Bocquet:11} --- made of or coated with hydrophobic materials, and characterized by small-scale roughness, either random or periodic. When these surfaces come into contact with liquid, air becomes trapped in a ``Cassie state'' within the cavities, creating a composite surface consisting of the solid substrate and the liquid-air menisci \citep{Quere:08}. Recently, a promising alternative has emerged: ``slippery liquid-infused porous surfaces'' (SLIPS), which are designed to spontaneously imbibe a liquid lubricant through wicking \citep{Lafuma:11,Wong:11,Hardt:2022}. While the liquid lubricant cannot match air’s low viscosity, SLIPS generally offer superior robustness compared with superhydrophobic surfaces. 

Towards characterizing the drag-reduction efficacy of a given microstructured surface, it is expedient to consider a canonical  problem \citep{Davis:09} where the liquid adjacent to the surface is exposed to a far-field simple shear flow, parallel to the surface, 
\begin{equation}\label{far field intro}
\bu_* \sim \hat{\mathbf{s}}\,G_*y_* \quad \text{as} \quad y_*\to\infty,
\end{equation}
where $y_*$ is the distance from the surface, $G_*$ is the shear rate of the imposed flow, and $\hat{\mathbf{s}}$ is its direction. (Henceforth, an asterisk decoration denotes a dimensional quantity.) Since the imposed flow \eqref{far field intro} vanishes at $y_*=0$, it constitutes an exact solution in the case of a smooth solid surface. The microstructure disturbs that flow, inducing a uniform ``slip'' stream, say $\bu^s_*$, at large distances whereby the far-field expansion \eqref{far field intro} extends as 
\begin{equation}\label{far field intro extended}
\bu_* = \hat{\mathbf{s}}\,G_*y_*+\bu^s_* + o(1) \quad \text{as} \quad y_*\to\infty. 
\end{equation}
With negligible inertia, the flow is governed by the linear Stokes equations. Assuming that the shape of the menisci is determined by capillarity (and, in particular, is essentially unaffected by the flow), the entire problem is linear and homogeneous in $\hat{\mathbf{s}}G_*$. The slip velocity must therefore possess the form $\bu^s_*=G_*\boldsymbol{\lambda}_*\bcdot\hat{\mathbf{s}}$, wherein the effective slip length $\boldsymbol{\lambda}_*$ is a two-dimensional (2D) second-order tensor depending on the surface geometry and the viscosity ratio of the two fluids. The most common microstructure configuration, used for both superhydrophobic surfaces and SLIPS, consists of a periodic array of ridges protruding perpendicularly from a flat solid substrate, with the lubricant filling the resulting grooves of rectangular cross section. For such configurations, the effective slip length adopts the simplified anisotropic form $\boldsymbol{\lambda}=\lambda^{\parallel}_*\be_3\be_3+\lambda^{\perp}_*\be_1\be_1$, in which the longitudinal slip length $\lambda^{\parallel}_*$ and transverse slip length $\lambda^{\perp}_*$ can be independently found by considering canonical flow problems where the shear is applied parallel ($\hat{\mathbf{s}}=\be_3$) or perpendicular to the ridges ($\hat{\mathbf{s}}=\be_1$), respectively.

Given the minute viscosity of air, the menisci of superhydrophobic surfaces have traditionally been modeled as free surfaces \citep{Batchelor:book}. The canonical effective-slip problem then reduces to a liquid-flow problem, decoupled from the flow inside the cavities. Accordingly, the microstructure geometry is manifested only through the partitioning of the liquid boundary into solid-liquid and gas-liquid interfaces. Assuming flat menisci, the imposed far-field stress is resisted only by viscous shear stresses at the solid-liquid interfaces. Hence, the limit where the solid fraction of the compound liquid interface vanishes is singular. In particular, the effective slip length diverges in that limit. The rate of this divergence --- scaling logarithmically with the solid fraction for grooved or meshed surfaces, or algebraically for post arrays --- has been elucidated through numerous analytical and asymptotic studies \citep{Philip:72:integral,Lauga:03,Ybert:07,Davis:09:mesh,Davis:10,Schnitzer:16,Schnitzer:17,Yariv:17:amplification,Schnitzer:18:Fakir,Schnitzer:18,Yariv:18}. While giant slip lengths have indeed been measured using small-solid-fraction superhydrophobic surfaces \citep{Choi:06,Lee:08}, the prospects of such surfaces are considered limited since the Cassie state inevitably collapses if the solid fraction is too low \citep{Extrand:04,Sbragaglia:07:spontaneous,Lee:16}.

Unlike in the superhydrophobic scenario, models of flow over SLIPS typically account for a finite lubricant viscosity. The effective-slip problem then involves the coupling of the exterior flow outside the cavities and the lubricant flow inside them. This coupling introduces a dependency not just on the viscosity ratio of the two fluids, but also on the interior geometry of the microstructure cavities; for ``rectangular''  grooves, the extra geometric parameter, in addition to the solid fraction, is the ridge height (i.e., groove depth). Very few analytical studies of the effective slip problem have included the effect of the interior fluid. The only systematic analysis we are aware of is that of \citet{Crowdy:17:Perturbation}, who derived analytical formulas for the effective slip length by perturbing about Philip's classical complex-variable solutions for an inviscid lubricant \citep{Philip:72}. Other investigators, including \citet{Schonecker:14} and \citet{Nizkaya:14}, have employed a mix of exact solutions and heuristic modeling assumptions, demonstrating good agreement with numerical simulations over a wide parameter range.

Having a finite lubricant viscosity regularizes the small-solid-fraction singularity. Indeed, the imposed shear is then resisted not just by the exterior solid boundaries, as in the superhydrophobic scenario, but also by the liquid-liquid interfaces. (Alternatively, one could interpret this as the imposed shear being resisted by both the exterior and interior solid boundaries.) Although the effective slip length no longer diverges as the solid fraction vanishes, it is still expected to increase in magnitude in that limit. Moreover, from a practical viewpoint, the SLIPS are stable even at very low solid fraction \citep{Smith:13}. This motivates theoretical analysis of idealized \emph{zero-solid-fraction} configurations. In particular, it is natural to consider the canonical rectangular-grooves geometry in the case of zero-thickness ridges --- henceforth referred to as a ``grating.'' Indeed, this geometry 
has been studied both in the transverse problem \citep{Hocking:76} and the longitudinal problem \citep{Bechert:89}. (The latter problem was actually studied for a single-phase flow, corresponding to the special case where the viscosities are matched.) This geometry has also been employed by \citet{Yan:24} in an analysis of transverse flow over a liquid-infused surface, with the effective slip length estimated using a heuristic approach similar to those of \citet{Schonecker:14} and \citet{Nizkaya:14}. 

For the zero-solid-fraction grating geometry, the imposed shear is entirely supported by the lubricant. Accordingly, this scenario introduces a new \emph{singular} limit in which the effective slip length diverges: that where the lubricant viscosity is vanishingly small. This previously unexplored singular limit is practically significant as it characterizes the maximum potential for drag reduction of SLIPS. As such, the motivation for studying it parallels that for the singular small-solid-fraction limit of the inviscid-lubricant scenario. In light of the above, the majority of our paper is concerned with an asymptotic analysis of the grating geometry in the singular limit of a nearly inviscid lubricant. We shall focus on the longitudinal grating problem, which is simpler and associated with larger slip lengths than its transverse counterpart. Similarly to analyses of the above-mentioned small-solid-fraction limit, we shall employ matched asymptotic expansions \citep{Hinch:book} in conjunction with conformal mappings \citep{Brown:book}. 

While the singular nature of the nearly-inviscid limit is evident, the corresponding scaling of the effective slip length is not. Let $\lambda_*$ denote this scaling, and $\mu=\mu_*^-/\mu_*^+$ the ratio of the interior-to-exterior viscosities. By an integral force balance, the imposed shear stress $\mu_*^+G_*$ must balance the average stress at the liquid-liquid interface. In the limit $\mu\ll1$, it is intuitive to anticipate that the exterior flow is dominated by a uniform slip stream, which in light of \eqref{far field intro extended} should be of magnitude $v_*=G_*\lambda_*$. Since both the velocity and shear stress are continuous at the interface, the latter can be estimated as $\mu^-_*v_*/\ell_*$, where $\ell_*$ is a length scale characterizing the interior flow --- perhaps the groove width, or its depth? This argument, which follows the classical scaling approach of \citet{Ybert:07}, yields the  singular scaling $\lambda_*/\ell_* = O(1/\mu)$. This scaling is built into Yan and Kowal's model for slip over a grating geometry \citep{Yan:24}. Some scrutiny reveals that it is also consistent with the more general models proposed by  \citet{Schonecker:14} and \citet{Nizkaya:14}, when these are  degenerated to a grating geometry and reduced in the nearly-inviscid limit. 

Nonetheless, our analysis will reveal that the $\mu^{-1}$ scaling is actually \emph{erroneous}, except in a lubrication-like limit corresponding to extremely shallow grooves. Indeed, the above scaling arguments overlook the formation of boundary layers, of \emph{exponentially small} extent, about the tips of the zero-thickness ridges. At leading order, the applied shear is entirely supported by the localized  flow in these tip regions. This delicate balance, in turn, implies an effective slip length that turns out to scale as $\mu^{-1/2}$. 

The paper is arranged as follows. In the next section we formulate the problem for longitudinal shear flow about a grating and derive a useful integral force balance. We also briefly discuss a numerical scheme used to solve this ``exact'' problem. The intuitive lubrication limit corresponding to extremely shallow grooves is briefly addressed in Sec.~\ref{sec:shallow}. In preparation for the analysis of the singular limit $\mu\ll1$, we perform in Sec.~\ref{sec:local} a local analysis of the flow problem in the vicinity of the tips. The limit $\mu\ll1$ is analyzed in Sec.~\ref{sec:Nearly inviscid}, where the identification and analysis of the localized tip layers furnishes the leading-order slip length. In Sec.~\ref{sec:slip 0}  we derive a functional expression for the $\ord(1)$ correction to the slip length. 
In Sec.~\ref{sec:dist} we identify a distinguished limit of shallow grooves and a nearly-inviscid lubricant and calculate the associated slip length. We conclude in Sec.~\ref{sec:conclude}, addressing deviations from a grating geometry to finite solid fractions.

\section{Problem formulation}
\begin{figure}[hbtp]
\centering
\includegraphics{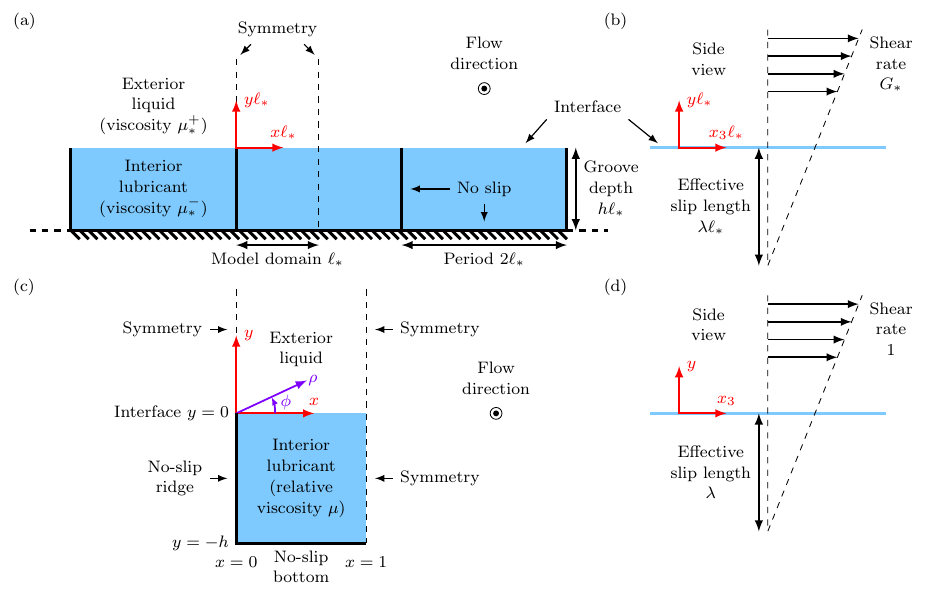}
\caption{Schematics showing (a,b) dimensional and (c,d) dimensionless quantities. In (c), only the model domain $0 < x < 1$ is shown.}
\label{fig:setup}
\end{figure}

The problem is depicted in Fig.~\ref{fig:setup}(a,b). The underlying solid substrate is a grating of period $2\ell_*$. The ridges are assumed infinitely thin; their height is $h \ell_*$. The grooves bounded by the ridges and the bottom substrate are filled with an ``interior'' lubricant, of viscosity $\mu^-_*$. The resulting compound surface is brought into contact with an ``exterior'' liquid of viscosity $\mu_*^+$. We assume that the menisci, pinned at the tips of the ridges, are flat (zero protrusion angle). At large distances from the surface, the flow approaches a longitudinal shear flow of shear-rate magnitude $G_*$; the velocity field thus satisfies the far-field condition \eqref{far field intro}, with the unit vector $\hat{\mathbf{s}}$ parallel to the grooves.

We utilize a dimensionless notation where length variables are normalized by $\ell_*$ and velocities by $\ell_* G_*$. 
We employ Cartesian coordinates $(x_1=x,x_2=y,x_3)$, with the imposed shear in the $x_3$-direction and the ridges located at $x=2n$ with integer $n$. The plane $y=0$ passes through the ridge tips; the external and internal fluids occupy the domains $y>0$ and $-h<y<0$, respectively. Since the flow is unidirectional, the velocity fields in the interior and exterior domains can be written as $\be_3 w^\pm(x,y)$, where $\pm$ indicates the sign of $y$. Due to the discrete-translational (periodic) and mirror symmetries, it suffices to focus upon a single ``semi-unit cell'', say, extending from the ridge at $x=0$ to the center of the groove at $x=1$, see  Fig.~\ref{fig:setup}(c).

The equations governing the velocity fields $w^\pm$ are then given as follows. Due to unidirectionality, the Navier--Stokes equations for the flows simplify to Laplace's equation in each domain,
\begin{equation}
\pd{^2w^\pm}{x^2}+\pd{^2w^\pm}{y^2} = 0 \quad \text{for} \quad y>0\ (+) \quad \text{or} \quad {-h} < y< 0\ (-). \label{Laplace both}
\end{equation}
The two fields are linked by the interfacial conditions of continuous velocity,
\begin{subequations}\label{continuity}
\begin{equation}
w^+=w^- \quad \text{at} \quad y=0, \label{continuous velocity}
\end{equation}
and continuous shear stress,
\begin{equation}\label{continuous stress}
\quad \pd{w^+}{y}=\mu \pd{w^-}y \quad \text{at} \quad y=0,
\end{equation}
\end{subequations}
where $\mu=\mu_*^-/\mu_*^+$. The interior field $w^-$ further satisfies the symmetry condition 
\begin{equation}\label{symmetry -}
\pd{w^-}{x} = 0  \quad \text{at} \quad x=1,
\end{equation}
and the no-slip condition at the ridge and at the bottom,
\begin{subequations}\label{noslip both}
\refstepcounter{equation}\label{noslip ridge}
\refstepcounter{equation}\label{noslip bottom}
\begin{equation}\tag{\ref*{noslip both}a,b}
w^-=0 \quad \text{at} \quad x=0, 
\qquad
w^-=0 \quad \text{at} \quad y=-h.
\end{equation}
\end{subequations}
The exterior field $w^+$ further satisfies the two symmetry conditions 
\begin{subequations}\label{symmetry+ both}
\refstepcounter{equation}\label{symmetry+ x=0}
\refstepcounter{equation}\label{symmetry+ x=1}
\begin{equation}\tag{\ref*{symmetry+ both}a,b}
\pd{w^+}{x} = 0  \quad \text{at} \quad x=0,
\qquad
\pd{w^+}{x} = 0  \quad \text{at} \quad x=1,
\end{equation}
\end{subequations}
the imposed shear in the far field,
\begin{equation}\label{far}
w^+\sim y \quad \text{as} \quad y\to\infty,
\end{equation}
and the no-slip condition at the tip of the ridge, 
\begin{equation}\label{tip}
\lim_{(x,y)\to(0,0)}w^+ = 0. 
\end{equation}

The problem posed by \eqref{Laplace both}--\eqref{tip} uniquely determines 
the flow fields $w^\pm(x,y)$, depending on the governing parameters $\mu$ and $h$. In particular, it determines the ``slip length'', 
\begin{equation}\label{lambda def}
    \lambda=\lim_{y\to\infty}(w^+-y),
\end{equation}
namely the constant correction to the shear in the far-field behavior \eqref{far}. That behavior is accordingly refined to [see Fig.~\ref{fig:setup}(d)]
\begin{equation}\label{far refined}
w^+ \os{ = } y + \lambda + o(1) \quad \text{as} \quad y\to\infty. 
\end{equation}
The goal is to calculate $\lambda$ as a function of $\mu$ and $h$. 

\newcommand{\Dtotal}{\mathcal{D}}
It is useful to define the integral quantity
\begin{subequations}\label{drag both}
\begin{equation}\label{drag ext}
\Dtotal = \int_{0}^1\left.\pd{w^+}{y}\right|_{y=0}\,\intd{x},
\end{equation}
which can be interpreted as the net viscous drag exerted by the exterior fluid on the interior fluid, through viscous shear stresses at the interface. Using \eqref{continuous stress}, we can also express this drag in terms of the interior field as
\begin{equation}\label{drag int} 
\Dtotal = \mu\int_{0}^1\left.\pd{w^-}{y}\right|_{y=0}\,\intd{x}.
\end{equation} 
\end{subequations}

An integral force balance can be used to derive alternative expressions for $\Dtotal$. Let $\mathcal{C}$ be any simple curve in the exterior domain extending from the line $x=0$ to the line $x=1$. By integrating Laplace's equation \eqref{Laplace both} over the domain bounded between $y=0$, the curve $\mathcal{C}$ and the lines $x=0,1$, and making use of the 2D variant of Gauss's theorem in conjunction with the symmetry conditions \eqref{symmetry+ both}, we find 
\begin{equation}\label{drag general}
\Dtotal = \int_{\mathcal{C}}\hat{\mathbf{n}}\bcdot\boldsymbol{\nabla} w^+\,dl,
\end{equation}
where $\hat{\mathbf{n}}$ is the unit normal to $\mathcal{C}$ pointing away from the domain, $\boldsymbol{\nabla}=\be_1\partial/\partial x+\be_2\partial/\partial y$ and $dl$ is the differential arc length along $\mathcal{C}$.

The viscous drag at the interface  must balance with the viscous traction at large distances from the surface, which is set by the external shear flow. Indeed, following Schnitzer \cite{Schnitzer:16,Schnitzer:17}, we shift  $\mathcal{C}$ to infinity in \eqref{drag general} and use the far-field condition \eqref{far} to find 
\begin{equation}\label{drag unity}
\Dtotal = 1. 
\end{equation}
Substituting this result into \eqref{drag both}--\eqref{drag general} gives the three equivalent integral balances 
\begin{subequations}\label{balances}
\refstepcounter{equation}\label{balance ext}
\refstepcounter{equation}\label{balance int}
\refstepcounter{equation}\label{balance general}
\begin{equation}\tag{\ref*{balances}a,b,c}
\int_{0}^1\left.\pd{w^+}{y}\right|_{y=0}\,\intd{x}=1, \quad \mu\int_{0}^1\left.\pd{w^-}{y}\right|_{y=0}\,\intd{x}=1, \quad \int_{\mathcal{C}}\hat{\mathbf{n}}\bcdot\boldsymbol{\nabla} w^+\,dl=1.
\end{equation}
\end{subequations}
As these balances follow from the problem formulation, they do not provide independent constraints. Nonetheless, they will be found to be useful in the subsequent analysis. In particular, \eqref{balance int} can be imposed on the interior field without the need to calculate the corresponding exterior solution. 

We can also derive an integral expression for the slip length in terms of the velocity distribution at the interface. We first note from \eqref{drag general} that the drag definition \eqref{drag ext} can be generalized to arbitrary $y\ge0$:
\begin{equation}\label{drag ext general}
\Dtotal = \int_0^1\pd{w^+}{y}\,dx. 
\end{equation}
Upon interchanging integration and differentiation, this extension gives  
\begin{equation}\label{drag all y}
\Dtotal = \frac{d}{dy}\int_0^1w^+(x,y)\,dx.
\end{equation}
Then, by integrating \eqref{drag all y} with respect to $y$, using \eqref{drag unity} and the refined far-field behavior \eqref{far refined}, and subsequently evaluating at $y=0$, we find 
\begin{equation}\label{slip integral}
\lambda=\int_0^1w^+(x,0)\,dy.
\end{equation}

Our main goal in this paper is to derive approximations for the slip length $\lambda$ in the limit $\mu\ll1$, corresponding to a nearly inviscid lubricant. Prior to this, in the following Sec.~\ref{sec:shallow}, we briefly consider the limit of shallow grooves, $h\ll1$, and, in the subsequent Sec.~\ref{sec:local}, we perform a local analysis of the exact problem near the ridge tip. These preliminary steps will help guide our analysis and physical interpretation of the results. 

\begin{figure}[t!]
\begin{center}
\includegraphics{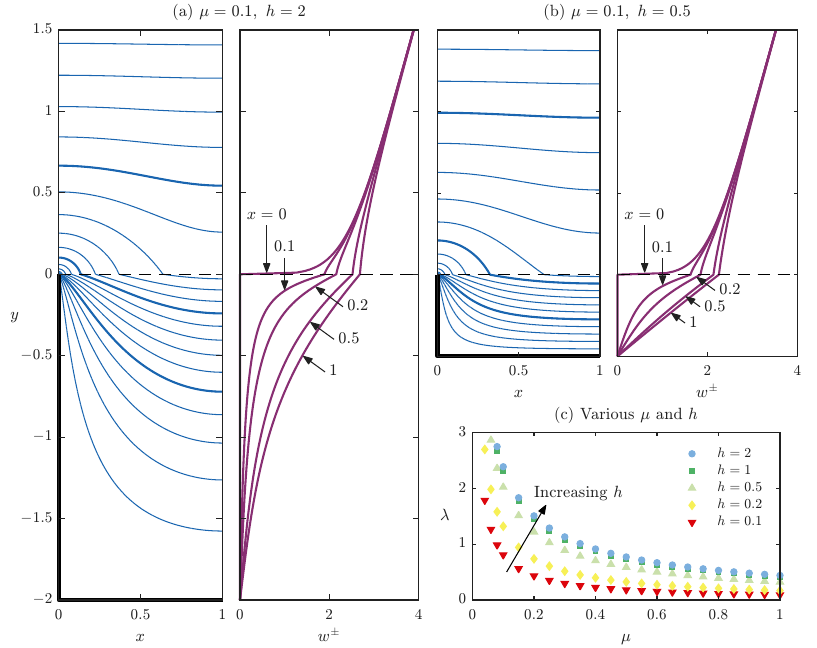}
\caption{Example numerical results. In (a) and (b), the left-hand panels show contour plots of $w^\pm(x,y)$ with contour spacing $\Delta w = 0.2$ (thin contours) and $\Delta w = 1$ (thick contours); the no-slip walls [$w^- = 0$, see~\eqref{noslip both}] are marked with thick black lines. The right-hand panels show the $w^\pm$ profile at various values of $x$. The interior-to-exterior viscosity ratio is $\mu = 0.1$, and the groove depth is (a) $h = 2$ (corresponding to grooves of square cross section) and (b) $h = 1/2$. Panel (c) shows the resulting slip length $\lambda$ as function of $\mu$, for various values of $h$.}
\label{fig:numerical}
\end{center}
\end{figure}

We have also implemented a numerical scheme (see Appendix~\ref{app:numerical} for details) to solve a boundary-integral formulation of the governing equations \eqref{Laplace both}--\eqref{tip}, with the slip length evaluated from its integral representation \eqref{slip integral}. Figure \ref{fig:numerical} presents sample velocity fields for $\mu = 0.1$ and two values of $h$ in detail, and the resulting slip length for various values of $\mu$ and $h$. The singularity of the limit $\mu\ll1$ is evident. Further results are presented throughout the paper as a comparison with the results of our asymptotic analysis.

\section{Shallow grooves} \label{sec:shallow}
We begin by briefly considering the limit of shallow grooves, $h\ll1$. It is readily seen that the one-dimensional linear profiles
\begin{subequations}\label{1D profiles}
\refstepcounter{equation}\label{1D profiles +}
\refstepcounter{equation}\label{1D profiles -}
\begin{equation}\tag{\ref*{1D profiles}a,b}
	w^+ = y + \frac{h}{\mu}, \qquad w^- = \frac{y+h}{\mu}, 
\end{equation}
\end{subequations}
satisfy the governing equations \eqref{Laplace both}--\eqref{tip}, except for the no-slip conditions \eqref{noslip ridge} and \eqref{tip} on the ridge. This suggests that \eqref{1D profiles} should be interpreted asymptotically as leading-order ``outer'' approximations in the limit $h\ll1$ with $y/h$ fixed \citep{Hinch:book}, which break down near the ridge. The definition \eqref{lambda def} then yields the slip length
\begin{equation}\label{slip shallow}
\lambda \approx \frac{h}{\mu}, 
\end{equation}
which is small (as long as $h \ll \mu$). This is to be expected, as the flow of the lubricant is resisted by the bottom boundary. 

While the lubrication-type approximation \eqref{slip shallow} is simple and intuitive, we shall see that its validity for small $\mu$ is severely limited.

\section{Local analysis near the ridge tip} \label{sec:local}
We next analyze the local structure of the velocity field in the vicinity of the ridge tip $(x,y)=(0,0)$. We start by reformulating the relevant governing equations in terms of polar coordinates $(\rho,\phi)$ centered about the tip, defined by $(x,y) = (\rho\cos\phi,\rho\sin\phi)$ [see Fig.~\ref{fig:setup}(c)]. 
These include Laplace's equation [cf.~\eqref{Laplace both}], 
\begin{equation}
\rho\pd{}{\rho}\left(\rho\pd{w^\pm}{\rho}\right) + \pd{^2w^\pm}{\phi^2}=0
 \quad \text{for} \quad 0<\pm\phi<\frac{\pi}{2}; \label{Laplace polar}
\end{equation}
the conditions of continuous velocity and shear stress [cf.~\eqref{continuity}], 
\begin{subequations}\label{continuity polar}
\refstepcounter{equation}\label{continuous velocity polar}
\refstepcounter{equation}\label{continuous stress polar}
\begin{equation}\tag{\ref*{continuity polar}a,b}
w^+=w^- \quad \text{at} \quad \phi=0, \qquad
\pd{w^+}{\phi}=\mu\pd{w^-}{\phi} \quad \text{at} \quad \phi=0;
\end{equation}
\end{subequations}
the requirements at $x=0$ for no slip on the ridge and symmetry in the exterior domain [cf.~\eqref{noslip ridge} and \eqref{symmetry+ x=0}],
\refstepcounter{equation}\label{x=0 polar}
\begin{equation}\tag{\theequation a,b}
w^-=0 \quad \text{at} \quad \phi = -\frac{\pi}{2}, \qquad
\pd{w^+}{\phi}=0 \quad \text{at} \quad \phi = \frac{\pi}{2};
\end{equation}
and the tip condition [cf.~\eqref{tip}],
\begin{equation}\label{tip polar}
\lim_{\rho\to0}w^+ = 0. 
\end{equation}
The remaining conditions [\eqref{symmetry -}, \eqref{noslip bottom}, \eqref{symmetry+ x=1} and \eqref{far}] do not affect the local analysis as they apply away from the tip. 

We now seek a leading-order approximation of the velocity field as $\rho\searrow0$. We denote this approximation by $w_L^\pm(\rho,\phi)$ and assume that it is a power law in $\rho$, i.e., proportional to $\rho^\alpha$,
where the unknown exponent $\alpha$ is positive in order to satisfy \eqref{tip polar}. The general power-law solution of \eqref{Laplace polar} is of the form
\begin{equation}\label{local ansatz}
	w_L^\pm(\rho,\phi) = \rho^\alpha\left[A^\pm \cos(\alpha \phi) + B^\pm \sin(\alpha\phi)\right].
\end{equation}
Continuity of velocity, \eqref{continuous velocity polar}, demands $A^+ = A^-$; we denote this common value by $A$. Continuity of shear stress, \eqref{continuous stress polar}, and the no-slip and symmetry conditions, \eqref{x=0 polar}, yield the relations
\begin{equation}
    B^+ = \mu B^-, \quad A \cos\frac{\alpha\pi}{2} - B^- \sin\frac{\alpha\pi}{2} = 0, \quad -A \sin \frac{\alpha \pi}{2} + B^+ \cos \frac{\alpha\pi}{2} = 0.
\end{equation}
Upon eliminating $B^\pm$, we obtain the condition $\tan (\alpha\pi/2) = \pm \mu^{1/2}$ determining the allowed values of $\alpha$. The dominant local behavior is given by the smallest (positive) value,
\begin{equation}\label{local alpha}
    \alpha = \frac{2}{\pi}\arctan \mu^{1/2}.
\end{equation}
The associated solution is
\refstepcounter{equation}\label{local w solution}
\begin{equation}\tag{\theequation a,b}
    w_L^+ = A \rho^\alpha \left[\cos(\alpha \phi) + \mu^{1/2} \sin(\alpha\phi)\right], \quad w_L^- = A\rho^\alpha \left[\cos(\alpha \phi) + \mu^{-1/2} \sin(\alpha\phi)\right].
\end{equation}
The constant $A$ is an overall amplitude that is determined by global considerations.

We conclude that, at the interface $y=0$, the leading-order local behavior of the velocity and shear stress are given by 
\refstepcounter{equation}\label{local interface}
\begin{equation}\tag{\ref*{local interface}a,b} 
	w_L^+ = w_L^- = Ax^\alpha, \qquad \pd{w_L^+}{y} = \mu \pd{w_L^-}{y} = A\alpha \mu^{1/2} x^{\alpha-1}.
\end{equation}
This local solution suggests that the contribution to the drag integral \eqref{drag both} from some small region $0< x< \delta$, where $\delta\ll1$, can be approximated as 
\begin{equation}\label{local drag}
\mu    \int_0^{\delta}  \left.\pd{w^-}{y}\right|_{y=0}\,\intd{x}  \approx  A \mu^{1/2} \delta^{\alpha}.
\end{equation}
We shall employ this estimate for scaling purposes.

\section{Nearly inviscid lubricant} \label{sec:Nearly inviscid}
We now consider the limit
\begin{equation}
\mu\ll1\label{mu<<1}
\end{equation}
where we anticipate that the slip length is asymptotically large. 

\subsection{Scaling arguments based on local analysis}\label{sec:Intuition}

While the local analysis of Sec.~\ref{sec:local} was derived in the limit $\rho \to 0$ for fixed $\mu$, it turns out that expanding its results for $\mu\ll1$ can give some intuition for the behavior of the solution near the tip when $\mu$ is small. For small $\mu$, we find that the exponent $\alpha$ of the local behavior, given by \eqref{local alpha}, is approximately
\begin{equation}\label{alpha approx}
    \alpha \approx \frac{2\mu^{1/2}}{\pi},
\end{equation}
which is also $\ll1$. %Since $\rho^\alpha = \ee^{\alpha\ln \rho}$, we can then (
Provided $\alpha\ln \rho$ is small, we can then approximate
\begin{equation}\label{local outer} 
    \rho^\alpha \approx 1 + \mu^{1/2}\frac{2}{\pi}\ln\rho.
\end{equation}
Thus, the leading-order local approximations 
\eqref{local w solution} simplify to
\refstepcounter{equation}\label{local small mu}
\begin{equation}\tag{\theequation a,b}
    w^+ \approx A\left(1 + \mu^{1/2} \frac{2}{\pi}\ln \rho\right), \quad w^- \approx A\left(1 + \frac{2\phi}{\pi}\right)\left(1 + \mu^{1/2} \ln \rho\right).
\end{equation}

We first note that the approximations \eqref{local small mu} explain the near-tip behavior seen in Fig.~\ref{fig:numerical}(a,b), where $\mu = 0.1$. In the exterior domain ($y>0$), the contours are quarter-circular due to the independence upon $\phi$, but in the interior domain ($y < 0$) the contours are radial due to the leading-order independence upon $\rho$. 

Secondly, the estimate \eqref{local drag} for the contribution to the drag integral \eqref{drag both} from a small region near the tip, using the small-$\mu$ approximation \eqref{alpha approx}, becomes 
\begin{equation}\label{local drag small mu}
\mu    \int_0^{\delta}  \left.\pd{w^-}{y}\right|_{y=0}\,\intd{x}  \approx  A \mu^{1/2}
\end{equation} 
(provided that $\mu^{1/2}\ln \delta$ is small). Assuming that this forms a non-negligible part of the total drag, \eqref{drag unity} suggests that $A = \ord(\mu^{-1/2})$. [In fact, it will turn out that this is the sole contribution to the drag integral at $\ord(1)$, so that the force balance \eqref{balance int} gives $A \sim \mu^{-1/2}$.]

\subsection{Asymptotic structure for nearly inviscid lubricant}\label{ssec:asymstructure}
For $\mu \ll 1$, the intuitive arguments in Sec.~\ref{sec:Intuition} suggest that velocities should be $\ord(\mu^{-1/2})$ in order to result in an $\ord(1)$ drag. We therefore postulate the expansions
\begin{align}\label{outer expansion w}
	w^\pm &\sim \mu^{-1/2} w_{-1}^\pm(x,y) + w_0^\pm(x,y) + \mu^{1/2} w_1^\pm(x,y) + \cdots,
	\\
	\label{expansion lambda}
	\lambda &\sim \mu^{-1/2} \lambda_{-1} + \lambda_0 + \mu^{1/2} \lambda_1 + \cdots,
\end{align}
as $\mu \to 0$. The far-field condition \eqref{far refined} then yields
\begin{subequations}\label{outer far}
\refstepcounter{equation}\label{outer far -1}
\refstepcounter{equation}\label{outer far 0}
\begin{equation}\tag{\ref*{outer far}a,b}
	w_{-1}^+ \sim \lambda_{-1}, \qquad w_0^+ \sim y + \lambda_0\os{,} 
\end{equation}
\end{subequations}
as $y \to \infty$, and continuity of shear stress \eqref{continuous stress} yields
\begin{subequations}\label{outer stress}
\refstepcounter{equation}\label{outer stress -1}
\refstepcounter{equation}\label{outer stress 0}
\begin{equation}\tag{\ref*{outer stress}a,b,c}
	\pd{w_{-1}^+}{y} = 0, \qquad \pd{w_0^+}{y} = 0, \qquad \pd{w_1^+}{y} = \pd{w_{-1}^-}{y}, 
\end{equation}
\end{subequations}
at $y = 0$. The remainder of the equations \eqref{Laplace both}--\eqref{tip} simply retain their homogeneous form at each order.

It may appear from (\ref{outer stress}a,b) and \eqref{drag both} that the drag $\mathcal{D}$ is $O(\mu^{1/2})$,  seemingly too small to satisfy \eqref{drag unity}. The resolution to this apparent contradiction is that the expansions \eqref{outer expansion w} represent an outer asymptotic region where $(x,y) = \ord(1)$, whereas the dominant force is provided by an inner asymptotic region about the tip of the ridge, $(x,y) = (0,0)$. (The scaling of the thickness of this inner region will be determined later.) The no-slip condition at the tip, \eqref{tip}, is accordingly omitted from the outer problems posed above, as it rather applies to the inner region. Similarly, the shear-stress-continuity conditions \eqref{outer stress} are henceforth understand to hold for $x>0$. 

An alternative (perhaps more tempting)  resolution would be to assume that the velocities and slip length are $\ord(\mu^{-1})$ instead, to directly yield an $\ord(1)$ shear stress at the interface. However, following such a calculation through, or appealing to \eqref{local drag small mu}, would reveal an $\ord(\mu^{-1/2})$ contribution from near the tip to the drag integral \eqref{drag both}, in contradiction to \eqref{drag unity}. Attempting any other large scaling for $w^\pm$ and $\lambda$ similarly results in a viscous drag of the wrong order; only the scaling \eqref{outer expansion w} yields a self-consistent result.

We now proceed to calculate the leading-order slip length $\lambda_{-1}$ by considering the outer approximation in the exterior domain, the inner approximation in both the exterior and interior domains, and the force balances  \eqref{balances}.

\subsection{Outer approximation in the exterior domain}\label{sec:uniform ext}

We first consider the leading-order outer flow in the exterior domain, $w_{-1}^+$. It satisfies Laplace's equation \eqref{Laplace both} together with the lateral symmetry conditions \eqref{symmetry+ both}, the far-field streaming  condition \eqref{outer far -1}, and the shear-free interface condition \eqref{outer stress -1}. This outer problem suggests the uniform flow
\begin{equation}\label{uniform ext}
w^+_{-1}\equiv \lambda_{-1}. 
\end{equation}

The leading-order outer approximation \eqref{uniform ext} fails to satisfy no slip at the ridge tip; this, of course, is in accordance with the supposed existence of the inner region. In order to identify the scaling of that region, it is helpful to briefly consider the outer correction $w_0^+$ --- in particular, its behavior as $\rho\to0$. (A detailed analysis of $w_0^+$, which is not necessary for determining the leading-order slip length, will be carried out in Sec.~\ref{sec:slip 0}.) 

At $\text{ord}(1)$, the generalized drag expression \eqref{drag ext general} gives 
\begin{equation}\label{drag 0 naive}
	\int_0^1\pd{w_0^+}{y}\,\intd{x} = 1.
\end{equation} 
While the exact expression \eqref{drag ext general} holds for all $y\ge0$, it only implies \eqref{drag 0 naive} for $y>0$, since the outer expansion \eqref{outer expansion w} fails at the origin $(x,y)=(0,0)$. The ambiguity of \eqref{drag 0 naive} at $y=0$, in conjunction with the integral apparently vanishing there owing to the shear-free condition \eqref{outer stress 0}, suggests that $w_0^+$ has a singularity at the origin.  

To characterize the singularity, we employ the generalized force balance \eqref{balance general} at $\text{ord}(1)$, 
\begin{equation}\label{balance general ord 1}
\int_{\mathcal{C}}\hat{\mathbf{n}}\boldsymbol{\cdot}\boldsymbol{\nabla}w_0^+\,dl = 1,
\end{equation}
with $\mathcal{C}$ the union of a quarter-circle of radius $0<\delta<1$ centered at the origin (within the exterior semi-unit cell) and the interval $x\in(\delta,1)$ on $y=0$. Given the stress-free interface condition \eqref{outer stress 0}, only the quarter-circle has a non-zero contribution. By letting $\delta\to0$, we see that 
the total force emanating from this ``corner'' is $1$. We therefore expect the singularity to be a source of strength $4$, i.e., 
\begin{equation}\label{source}
w^+_0 \sim \frac{2}{\pi}\ln\rho \quad \text{as} \quad \rho  \to 0. 
\end{equation}
In principle, this asymptotic behavior could include more singular terms (without modifying the net force); these, however, can be ruled out via matching considerations. 

\subsection{Inner tip region}\label{sec:tip solution}

To determine the scaling of the inner region near the tip, we consider the local behavior as $\rho \to 0$ of the first two terms in the expansion \eqref{outer expansion w} for $w^+$. Equations \eqref{uniform ext} and \eqref{source} show that the expansion breaks down once the exponentially small scale set by $\ln(1/\rho) = \ord(\mu^{-1/2})$ is reached; on that scale, the second term becomes asymptotically comparable to the first. 

This observation suggests that an appropriate rescaled coordinate $\sigma$ for the inner region near the tip is given by
\begin{equation}\label{varrho def}
	\sigma = \mu^{1/2}\ln \rho.
\end{equation}
The coordinate $\sigma$ is negative, with the tip $\rho = 0$ located at $\sigma = -\infty$. Matching to the outer region $\rho = \ord(1)$ takes place for small $\sigma$, specifically on the intermediate scale $\mu^{1/2} \ll \sigma \ll 1$.

We define the tip-scale velocity fields $\omega^\pm(\sigma,\phi) = w^\pm(\rho,\phi)$. They satisfy the following equations [cf.~\eqref{Laplace polar}--\eqref{tip polar} in the local analysis]:
\addtocounter{equation}{1}%
\begin{subequations}\label{continuity tip}%
\refstepcounter{equation}\label{continuous velocity tip}%
\refstepcounter{equation}\label{continuous stress tip}%
\end{subequations}%
\begin{subequations}\label{x=0 tip}%
\refstepcounter{equation}\label{x=0 tip interior}%
\refstepcounter{equation}\label{x=0 tip exterior}%
\end{subequations}%
\addtocounter{equation}{-3}%
\begin{gather}
\label{Laplace tip}
\pd{^2\omega^{\pm}}{\phi^2}+\mu\pd{^2\omega^{\pm}}{\sigma^2}=0
 \quad \text{for} \quad 0<\pm\phi<\frac{\pi}{2},
\\
\refstepcounter{equation}\tag{\theequation a,b}
\omega^+=\omega^- \quad \text{at} \quad \phi = 0, \qquad \pd{\omega^+}{\phi}=\mu\pd{\omega^-}{\phi} \quad \text{at} \quad \phi=0,
\\
\refstepcounter{equation}\tag{\theequation a,b}
\omega^-=0 \quad \text{at} \quad \phi = -\frac{\pi}{2}, \qquad
\pd{\omega^+}{\phi}=0 \quad \text{at} \quad \phi = \frac{\pi}{2},
\\
\label{tip tip}
\lim_{\sigma\to-\infty}\omega^+ = 0.
\end{gather}

We accordingly posit the inner expansions
\begin{equation}\label{tip expansions}
 \omega^\pm(\sigma,\phi) \sim \mu^{-1/2}\omega_{-1}^\pm(\sigma,\phi) + \omega_0^\pm(\sigma,\phi) + \mu^{1/2} \omega_1^\pm(\sigma,\phi) + \cdots \quad \text{as} \quad \mu \to 0,
\end{equation}
and seek to calculate $\omega_{-1}^\pm$. 
The set of homogeneous equations  \eqref{Laplace tip}--\eqref{tip tip} is supplemented at each order with the requirement of matching to the outer region. 
In particular, at leading order, the uniform solution \eqref{uniform ext} implies the inhomogeneous matching condition
\begin{equation}\label{trivial match}
	\lim_{\sigma\to0}\omega^+_{-1} = \lambda_{-1}. 
\end{equation}

We start with the exterior domain.
It is easy to see from the governing equation \eqref{Laplace tip} with the symmetry condition \eqref{x=0 tip exterior} at $\ord(\mu^{-1/2})$ 
that $\omega^+_{-1}$ 
cannot depend upon $\phi$: 
\refstepcounter{equation}
\begin{equation}\tag{\theequation}
    \omega_{-1}^+ = \omega_{-1}^+(\sigma). 
\end{equation}
Thus, the governing equation \eqref{Laplace tip} at $\ord(\mu^{1/2})$ becomes 
\begin{equation}
\pd{^2\omega_1^+}{\phi^2} = -\frac{d^2\omega_{-1}^+}{d\sigma^2}.
\end{equation}
Integrating this over $0 \leq \phi \leq \pi/2$ and using 
the symmetry condition \eqref{x=0 tip exterior} yields
\begin{equation}\label{ext shear}
\pd{\omega^+_1}{\phi} = \frac{\pi}{2} \frac{d^2\omega_{-1}^+}{d\sigma^2} \quad \text{at} \quad \phi=0. 
\end{equation}

Consider now the interior domain at $\ord(\mu^{-1/2})$. Solving the governing equation \eqref{Laplace tip} 
with the no-slip condition \eqref{x=0 tip interior} and continuity condition \eqref{continuous velocity tip} gives
\begin{equation}
\omega^-_{-1}(\sigma,\phi)=\left(1 + \frac{2\phi}{\pi}\right)\omega^+_{-1}(\sigma).
\label{tip interior -1}
\end{equation}
Substitution of \eqref{ext shear} and \eqref{tip interior -1} into the shear-stress condition \eqref{continuous stress tip} at $\ord(\mu^{1/2})$ then gives an ordinary differential equation for $\omega_{-1}^+$,
\begin{equation}
\frac{d^2\omega_{-1}^+}{d\sigma^2} = \frac{4}{\pi^2}\omega_{-1}^+.
\end{equation}
The solution that satisfies the no-slip tip condition \eqref{tip tip} and matching condition \eqref{trivial match} is
\begin{subequations}\label{omega lo}
\begin{equation}\label{omega+ lo}
\omega^+_{-1}=\lambda_{-1}\ee^{2\sigma/\pi}.
\end{equation}
The corresponding  leading-order interior solution \eqref{tip interior -1} is 
\begin{equation}\label{omega- lo}
\omega^-_{-1}=\lambda_{-1}\left(1 + \frac{2\phi}{\pi}\right)\ee^{2\sigma/\pi}. 
\end{equation}
\end{subequations}
[Reassuringly, the form of the inner solutions \eqref{omega lo} agrees with the small-$\mu$ expansion \eqref{local small mu} of the results from the local analysis.]

\subsection{Leading-order drag and slip length}\label{sec:lo drag}
It remains to determine the slip length $\lambda_{-1}$. We turn to the viscous-drag integral \eqref{balance int}, and split it into separate contributions from the tip and outer regions, $\Dtotal=\Dtotal_\mathrm{tip}+\Dtotal_\mathrm{outer}$, where  
\newcommand{\Dtip}{\Dtotal_\mathrm{tip}}
\newcommand{\Douter}{\Dtotal_\mathrm{outer}}
\begin{subequations}\label{balance split}
\refstepcounter{equation}\label{balance split tip}
\refstepcounter{equation}\label{balance split outer}
\begin{equation}\tag{\ref*{balance split}a,b}
	\Dtip = \mu \int_0^\delta  \left.\pd{w^-}{y}\right|_{y=0}\,\intd{x}, \qquad \Douter = \mu\int_\delta^1 \left.\pd{w^-}{y}\right|_{y=0}\,\intd{x}\os{.}
\end{equation}
By virtue of the force balance \eqref{balance int},  
\begin{equation}\label{balance split sum}
\Dtip + \Douter =1.
\end{equation}
\end{subequations}
The splitting point $x = \delta$ is chosen to be on an ``intermediate'' scale between the outer scale $x = \ord(1)$ and the inner scale $\sigma = \ord(1)$ [see \eqref{varrho def}]. This intermediate scale  is represented by the conditions
\begin{subequations}\label{delta range}
\refstepcounter{equation}\label{delta range upper}
\refstepcounter{equation}\label{delta range lower}
\begin{equation}\tag{\ref*{delta range}a,b}
	\delta \ll 1 \quad \text{ and } \quad \ln\frac{1}{\delta} \ll \mu^{-1/2}.
\end{equation}
\end{subequations}
Since the outer solution is $w^- = \ord(\mu^{-1/2})$ [cf.~\eqref{outer expansion w}], the outer contribution $\Douter$ scales as $\mu^{1/2}$ and does not contribute at leading order to \eqref{balance split sum}. 

For the tip contribution, we transform via polar coordinates $(\rho,\phi)$ into tip coordinates $(\sigma,\phi)$ using \eqref{varrho def}, and obtain
\begin{equation}\label{lo tip contribution}
 \Dtip = %\mu\int_{0}^\delta \frac{1}{\rho} \left.\pd{w^-}{\phi}\right|_{\phi=0}\,\intd{\rho} = 
 \mu^{1/2} \int_{-\infty}^{-\mu^{1/2}\ln(1/\delta)}\left.\pd{\omega^-}{\phi}\right|_{\phi=0}\,\intd{\sigma}. 
\end{equation}
Using the leading-order tip solution \eqref{omega- lo} and replacing the upper integration limit with $0$ [due to \eqref{delta range lower}], we obtain the result
\begin{equation}\label{lo tip contribution result}
    \Dtip \sim \frac{2\lambda_{-1}}{\pi}\int_{-\infty}^0e^{2\sigma/\pi}\,d\sigma = \lambda_{-1}.
\end{equation}
We conclude from the force balance \eqref{balance split sum} that the leading-order slip length is simply given by
\begin{equation}\label{slip -1}
	\lambda_{-1} = 1,
\end{equation}
that is,
\begin{equation}\label{lo result}
    \lambda \sim \mu^{-1/2} \quad \text{for} \quad \mu \ll 1.
\end{equation}
[An alternative derivation makes use of van Dyke matching \citep{vanDyke:book}: Writing \eqref{omega+ lo} for $\omega_{-1}^+$ in terms of the outer coordinate $\rho$ and expanding for $\mu\ll1$ gives
\begin{equation}\label{interior tip lo in rho}
\mu^{-1/2} \lambda_{-1} + \frac{2\lambda_{-1}}{\pi}\ln\rho + \cdots.
\end{equation}
Comparison with \eqref{source} then yields the result \eqref{slip -1}.]

The leading-order results for $\mu\ll1$ can be summarized as follows. The velocity in the exterior domain $y>0$ is approximately uniform, $w^+ \approx \lambda$ [see \eqref{uniform ext}]. Given the need to satisfy the no-slip condition, this uniform ``slip flow'' is disturbed in an exponentially small inner region about the tip, see  \eqref{omega lo}. This disturbance, in turn, results in a viscous drag $\approx \mu^{1/2} \lambda$. Other contributions to the drag only enter at higher order, so the balance \eqref{balance ext} between the viscous drag and the unity imposed shear yields the result $\lambda \approx  \mu^{-1/2}$ [see \eqref{lo result}].

\section{First correction to slip length} \label{sec:slip 0}
In the previous section, we determined the first term in the expansion
\begin{equation}\label{lam 2-term setup}
\lambda \sim \mu^{-1/2} + \lambda_0 \quad \text{for} \quad \mu\ll1.
\end{equation}
We proceed here to determine the correction, $\lambda_0$, which depends upon the groove depth $h$. This involves calculating the first corrections (smaller by an order $\mu^{1/2}$) to the outer approximation in the exterior domain and the inner (tip) approximation in both domains, 
as well as a leading-order outer approximation in the interior domain. We subsequently use these results to extend  expansion \eqref{lo tip contribution result} for $\Dtip$ to higher order, and to calculate the leading-order contribution to $\Douter$ from the interior solution. 

\subsection{First correction to the outer approximation in the exterior domain} 
We begin with the asymptotic correction $w_0^+$ to the uniform leading-order outer approximation in the exterior domain, \eqref{uniform ext}. To aid the calculation, we introduce a shifted field,
\begin{equation}\label{w0 shift}
	\tilde{w}_0^+ = w_0^+ - \lambda_0,
\end{equation}
so that $\lambda_0$ is eliminated from the far-field condition \eqref{outer far 0}: 
\begin{equation}\label{refined far}
    \tilde{w}_0^+ = y + o(1) \quad \text{as} \quad y\to\infty.
\end{equation}

The problem governing $\tilde{w}_0^+$ is given by Laplace's equation [recall~\eqref{Laplace both}]; homogeneous Neumann conditions at $x=0$, $x=1$ and $y=0$, excluding the origin [cf.~\eqref{symmetry+ both} and \eqref{outer stress 0}]; the far-field condition \eqref{refined far}; and \eqref{source}, which follows from matching to the tip region [see \eqref{interior tip lo in rho}]. Hence, $\tilde{w}_0^+$ has a unique solution that is \emph{universal}, i.e., does not depend on any parameters. In particular, it follows from the local behavior \eqref{source} that we can define the pure number
\begin{equation}\label{def beta}
    \beta=\lim_{\rho\to0}\left(\tilde{w}_0^+-\frac{2}{\pi}\ln \rho\right), 
\end{equation}
and deduce that
\begin{equation}\label{source refined}
    w_0^+ \sim \frac{2}{\pi}\ln\rho + \lambda_0 + \beta \quad \text{as} \quad \rho\to0.
\end{equation}
Expression \eqref{source refined} needs to match the inner region at the next order. 

The universal problem possesses the explicit solution
(see Appendix~\ref{app:w0tilde+})
\begin{equation}\label{w0tilde+ main result}
\tilde{w}_0^+ = \frac{1}{\pi}\ln\left[4\left(\sin^2\frac{\pi x}{2}+\sinh^2\frac{\pi y}{2}\right)\right],
\end{equation}
whereby \eqref{def beta} gives 
\begin{equation}\label{beta}
\beta=\frac{2}{\pi}\ln\pi. 
\end{equation}

\subsection{First corrections to the inner tip approximations} \label{sec:outer int 0}
Continuing to the tip region, we observe that the analysis in Sec.~\ref{sec:tip solution} can be repeated at one higher order, to yield 
\begin{subequations}\label{higher order tip C0}
\refstepcounter{equation}\label{higher order tip C0 +}
\refstepcounter{equation}\label{higher order tip C0 -}
\begin{equation}
\omega^+_0=C_0\ee^{2\sigma/\pi}, \qquad
\omega^-_0=C_0\left(1 + \frac{2\phi}{\pi}\right)\ee^{2\sigma/\pi}.
  \tag{\ref*{higher  order tip C0}a,b}
\end{equation}
\end{subequations}
The constant $C_0$ is determined by 2--2 van Dyke matching \citep{vanDyke:book} to the outer exterior solution, as follows. Taking the inner field 
$\omega^+$ up to $\ord(1)$ [i.e., \eqref{omega+ lo} and \eqref{higher order tip C0 +}, making use of \eqref{slip -1}], rewriting in terms of the outer coordinate $\rho$ and expanding for small $\mu$ gives, when retaining terms up to $\ord(1)$,
\begin{equation}
    \mu^{-1/2} + \frac{2}{\pi}\ln\rho + C_0. \label{2--2}
\end{equation}
Comparing with the outer behavior [i.e., \eqref{uniform ext} and \eqref{source refined}], we conclude that $C_0 = \beta + \lambda_0$, and hence \eqref{higher order tip C0} becomes
\begin{subequations}\label{higher order tip}
\refstepcounter{equation}\label{higher order tip +}
\refstepcounter{equation}\label{higher order tip -}
\begin{equation}\tag{\ref*{higher order tip}a,b}
\omega^+_0=(\beta+\lambda_0)\ee^{2\sigma/\pi}, \qquad
\omega^-_0=(\beta+\lambda_0)\left(1 + \frac{2\phi}{\pi}\right)\ee^{2\sigma/\pi}.
\end{equation}
\end{subequations}
To complete the calculation and determine $\lambda_0$, we need to revisit the viscous-drag integral \eqref{balance split}.

\subsection{Corrections to the viscous drag}
We first evaluate the tip contribution $\Dtip$ in \eqref{balance split tip} to $\ord(\mu^{1/2})$ using the first two terms in the expansion \eqref{tip expansions}. The leading-order approximation \eqref{omega- lo} contributes, using \eqref{slip -1},
\begin{equation}
    \int_{-\infty}^{\mu^{1/2}\ln\delta} \frac{2}{\pi}\ee^{2\sigma/\pi}\,\intd{\sigma} \sim 1 + \mu^{1/2} \frac{2}{\pi}\ln\delta,
\end{equation}
where we again exploit the fact that $\mu^{1/2}\ln\delta \ll 1$, see \eqref{delta range}. Similarly, the correction \eqref{higher order tip -} contributes
\begin{equation}
    \mu^{1/2} \int_{-\infty}^{\mu^{1/2}\ln\delta}(\beta+\lambda_0) \frac{2}{\pi}\ee^{2\sigma/\pi}\,\intd{\sigma} \sim \mu^{1/2} (\beta + \lambda_0).
\end{equation}
Hence, the tip contribution is
\begin{equation}\label{Stip correction}
	\Dtip \sim 1 + \mu^{1/2} \left(-\frac{2}{\pi}\ln \frac{1}{\delta} + \beta + \lambda_0\right).
\end{equation}

The outer contribution $\Douter$ in \eqref{balance split outer} depends on the leading-order outer approximation in the interior domain, $w_{-1}^-$. Prior to calculating $w_{-1}^-$, we derive expressions for $\Douter$ and the correction $\lambda_0$ in terms of integral properties of that field. %can be expressed in terms of an  Prior to calculating this approximation, we determine an expression for the desired correction $\lambda_0$ in terms of $w_{-1}^-$. 

Given that $w_{-1}^-$ matches to the tip solution $\omega_{-1}^-$ [given by \eqref{omega- lo} and \eqref{slip -1}], we can deduce its local behavior
\begin{equation}
	w_{-1}^- \sim 1 + \frac{2\phi}{\pi} \quad \text{as} \quad \rho \to 0. 
\end{equation}
Consequently, the shear at the interface has the near-tip behavior
\begin{equation}\label{shear singularity}
    \left.\pd{w_{-1}^-}{y}\right|_{y=0} \sim \frac{2}{\pi x} \quad \text{as} \quad x \to 0,
\end{equation}
which results in a logarithmic divergence in the drag integral \eqref{balance split outer} as $\delta \to 0$. We subtract off this singularity from the integral to obtain
\begin{equation}\label{Souter regularized pre}
    \Douter \sim \mu^{1/2} \left[\int_\delta^1 \left(\left.\pd{w_{-1}^-}{y}\right|_{y=0} - \frac{2}{\pi x}\right)\,\intd{x} - \frac{2}{\pi}\ln\delta \right].
\end{equation}
This motivates defining a ``regularized'' drag contribution
\begin{equation}\label{F def}
	\mathcal{F}(h) = \int_0^1 \left(\left.\pd{w_{-1}^-}{y}\right|_{y=0} - \frac{2}{\pi x}\right)\,\intd{x},
\end{equation}
where we make it explicit that $\mathcal{F}$ depends only on the groove depth $h$. Equation \eqref{Souter regularized pre} then simplifies to
\begin{equation}\label{Souter}
	\Douter \sim \mu^{1/2} \left[\mathcal{F}(h) + \frac{2}{\pi}\ln\frac{1}{\delta}\right].
\end{equation}

Upon substitution of the drag contributions \eqref{Stip correction} and \eqref{Souter} into the force balance \eqref{balance split sum}, the terms involving $\ln \delta$ cancel out, as they should, and the remaining terms at $\ord(\mu^{1/2})$ determine the correction to the slip length,
\begin{equation}\label{slip 0 pre}
	\lambda_0 = -\mathcal{F}(h) - \beta,
\end{equation}
where $\beta$ is given by \eqref{beta}.

\subsection{Leading-order outer approximation in the interior domain}\label{sec:interior}
At this stage all that remains is to calculate the leading-order outer approximation in the interior domain, 
$w^-_{-1}$, and use \eqref{F def} and \eqref{slip 0 pre} to calculate $\lambda_0$. 

The field $w^-_{-1}$ satisfies Laplace's equation [cf.~\eqref{Laplace both}], no-slip conditions at $x=0$ and $y=-h$ [cf.~\eqref{noslip both}], and the symmetry condition \eqref{symmetry -} at $x=1$. The flow is driven by the leading-order uniform exterior flow $w_{-1}^+ \equiv 1$ [cf.~\eqref{uniform ext} and \eqref{slip -1}] via continuity of velocity at the interface \eqref{continuous velocity}, which yields
\begin{equation}\label{interfacial'}
w^-_{-1} = 1 \quad \text{at} \quad y=0.
\end{equation}
[The field $w_{-1}^-$ must also satisfy a matching condition to the inner solution \eqref{omega- lo} near the origin. A local analysis near the origin reveals, however,  that this condition is trivially satisfied provided $w_{-1}^-$ remains bounded.] We thus have a well-posed problem governing $w^-_{-1}$ for all values of the depth $h$. 

We first consider the limiting case of infinitely deep grooves, $h=\infty$, for which the no-slip condition at $y = -h$ is replaced by the decay condition
\begin{equation}\label{decay interior}
\lim_{y\to-\infty} w^-_{-1} = 0.
\end{equation}
The associated calculation of $w^-_{-1}$ is carried out in Appendix~\ref{app:w-1- deep}. It results in  
\begin{equation}\label{interior shear -1}
\left.\pd{w^-_{-1}}{y}\right|_{y=0} = \frac{1}{\sin(\pi x/2)}, 
\end{equation}
which indeed blows up as $x \to 0$ according to \eqref{shear singularity}. Substitution of \eqref{interior shear -1} into \eqref{F def} yields
\begin{equation}\label{quadrature inf h}
\mathcal F(\infty) = \frac{2}{\pi}\ln\frac{4}{\pi},
\end{equation}
whereby, from \eqref{slip 0 pre},
\begin{equation}\label{lam0 inf h} 
\lim_{h\to\infty}\lambda_0=-\frac{2}{\pi}\ln 4. 
\end{equation}
Thus, for $h = \infty$, \eqref{lo result} is refined to
\begin{equation}\label{two-term slip for inf deep}
\lambda \sim \mu^{-1/2} - \frac{2}{\pi}\ln4 \quad \text{for} \quad \mu\ll1.
\end{equation}

\begin{figure}[t!]
\begin{center}
\includegraphics[width=\textwidth]{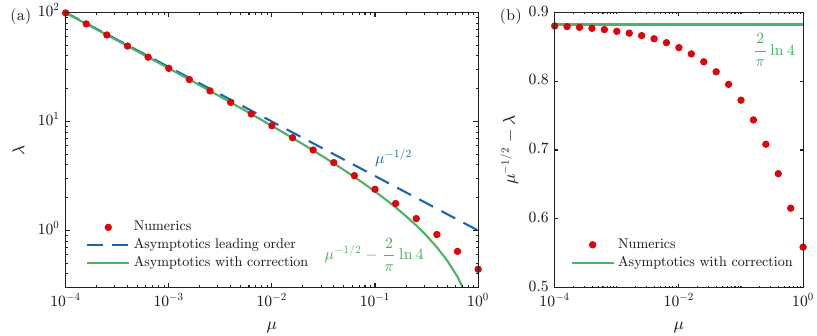}
\caption{(a) Slip length $\lambda$ as a function of viscosity ratio $\mu$ for deep grooves ($h = \infty$), comparing (symbols) numerical results with (dashed curve) the 
leading-order approximation \eqref{lo result} and (solid curve) the two-term approximation \eqref{two-term slip for inf deep}. (b) Deviation from the leading-order asymptotic prediction, comparing numerical results and the asymptotic prediction $-\lambda_0$ obtained from \eqref{lam0 inf h}.}
\label{fig:lambda_deep}
\end{center}
\end{figure}

In Fig.~\ref{fig:lambda_deep}(a), we compare the leading-order approximation \eqref{lo result} and two-term approximation \eqref{two-term slip for inf deep} with the slip length $\lambda$ calculated using the numerical scheme. Figure \ref{fig:lambda_deep}(b) shows the variation with $\mu$ of the difference $\mu^{-1/2} - \lambda$; it is evident that this difference indeed approaches \eqref{lam0 inf h}.

Another limit of interest is that of shallow grooves, $h\ll1$. Here the approximate solution $w^-_{-1}\sim 1 + y / h$ [analogous to \eqref{1D profiles -}] is readily obtained by disregarding the no-slip ridge condition \eqref{noslip ridge}. At leading order in $h$ we then find, from \eqref{F def} and \eqref{slip 0 pre}, 
\begin{equation}\label{slip0 h<<1 pre}
\lambda_0 \sim -\frac{1}{h} \quad \text{for} \quad h\ll1. 
\end{equation}

In principle, asymptotic corrections to 
\eqref{lam0 inf h} and \eqref{slip0 h<<1 pre} 
could be derived by perturbative analyses of the interior problem governing $w_{-1}^-$. In particular, the correction to the shallow-grooves approximation \eqref{slip0 h<<1 pre} could be obtained by analyzing the end region near $x=0$ \citep{Cormack:74,Sen:82} and its contribution to the integral \eqref{F def}. It turns out, however, that the problem for $w_{-1}^-$ can be solved exactly. 

The solution for arbitrary $h$ is derived in Appendix~\ref{sec:w-1- arb h} using a Schwarz--Christoffel mapping. It results in the regularized shear force
\begin{equation}\label{F arb h}
\mathcal F(h) = \frac{2}{\pi} \ln\frac{2}{\sqrt{1-k^2}\,K(k^2)}, 
\end{equation}
where $K$ is the complete elliptic integral of the first kind and $k$ is given by the implicit relation
\begin{equation}\label{k def}
\frac{K(1-k^2)}{K(k^2)}=h. 
\end{equation}
(We employ the convention where the argument of $K(m)$ is the ``parameter'' --- the square of the ``modulus,'' here corresponding to $k$.) 
Substitution of \eqref{F arb h} into \eqref{slip 0 pre} finally yields
\begin{equation}\label{lam0 arb h}
\lambda_0=\frac{2}{\pi} \ln\frac{\sqrt{1-k^2}\,K(k^2)}{2\pi},
\end{equation}
and hence the two-term approximation for the slip length, valid for arbitrary (fixed) $h>0$, is
\begin{equation}\label{lam 2-term result}
    \lambda \sim \mu^{-1/2} + \frac{2}{\pi} \ln\frac{\sqrt{1-k^2}\,K(k^2)}{2\pi} \quad \text{as} \quad \mu \to 0,
\end{equation}
where $k$ depends on $h$ via \eqref{k def}.

Revisiting the limits of deep and shallow grooves, we can expand \eqref{lam0 arb h} using the asymptotic behavior of the elliptic integral \citep{Abramowitz:book}. For deep grooves, where $k\searrow0$, we obtain
\begin{equation}\label{lam0 large h}
\lambda_0=-\frac{2}{\pi}\ln 4 - \frac{8}{\pi}\ee^{-\pi h} + O(\ee^{-2\pi h})\quad \text{for} \quad h\gg1
\end{equation} 
which provides a refinement of \eqref{lam0 inf h}. 
For shallow grooves, where $k\nearrow1$, we obtain
\begin{equation}\label{lam0 small h}
\lambda_0 = -\frac{1}{h}+\frac{2}{\pi}\ln\frac{1}{h} +O(\ee^{-\pi/h})\quad \text{for} \quad h\ll1,
\end{equation}
which provides a refinement of \eqref{slip0 h<<1 pre}.

\begin{figure}[t!]
\begin{center}
\includegraphics[width=\textwidth]{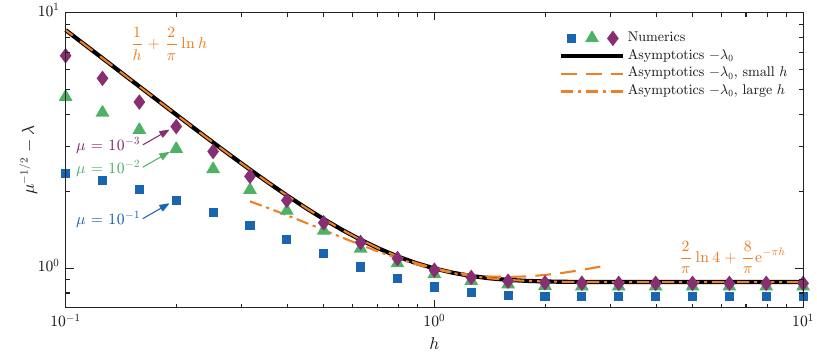}
\caption{Variation with groove depth $h$ of $\mu^{-1/2} - \lambda$, the (negative of the) slip-length deviation from the leading-order asymptotic prediction (cf.~Fig.~\ref{fig:lambda_deep}). The symbols show numerical results for three values of the viscosity ratio $\mu$. The curves show asymptotic results for $-\lambda_0$: solid, exact result \eqref{lam0 arb h}; dashed, small-$h$ approximation \eqref{lam0 small h}; and dash-dotted, large-$h$ approximation \eqref{lam0 large h}.}
\label{fig:lambda0}
\end{center}
\end{figure}

The variation with $h$ of $-\lambda_0$ is shown in Fig.~\ref{fig:lambda0}. Unsurprisingly, the agreement between the asymptotic and numerical results improves as $\mu$ diminishes. We note, however, that the asymptotic scheme fails as $h$ becomes small, even for small values of $\mu$. This failure motivates a careful look at the scenario where both $\mu$ and $h$ are small. 

\section{Shallow grooves: Distinguished limit and slip length} \label{sec:dist}

According to \eqref{lam0 small h}, the correction $\lambda_0$ in the slip-length expansion \eqref{lam 2-term setup} grows in magnitude as $h$ decreases. Hence, the expansion breaks down for $h$ on the order of $\mu^{1/2}$, when the correction becomes comparable to the leading-order term. Indeed, a breakdown for sufficiently small $h$ may be anticipated given the need to transition to the shallow-groove approximation \eqref{slip shallow}. 

A leading-order approximation describing that transition can be derived intuitively following  arguments analogous to those employed at the end of Sec.~\ref{sec:lo drag}. As before, the leading-order uniform exterior solution $w^+ \approx \lambda$ [cf.~\eqref{uniform ext}] drives 
shear near the tip which contributes a drag $\Dtip \approx \mu^{1/2}\lambda$ to the force integral \eqref{balance split} [see~\eqref{lo tip contribution result}]. The exterior solution also drives flow in the interior fluid, which due to the shallow geometry is approximately $w^- \approx \lambda(1 + y/h)$ [cf.~\eqref{1D profiles -}]. This  contributes a drag $\Douter \approx \mu \lambda/h$, which for $h$ small as $\mu^{1/2}$ becomes comparable to $\Dtip$. 
Since the total drag adds up to unity [see~\eqref{balance split sum}], we can solve for $\lambda$ to obtain the result
\begin{equation}\label{dist intuitive slip}
    \lambda \approx \frac{1}{\mu^{1/2} + \mu/h} 
    \quad \text{for } \mu\ll1, 
\end{equation}
with the two terms in the denominator being comparable for $h$ on the order of $\mu^{1/2}$. 
If $h \gg \mu^{1/2}$, $\Douter$ is negligible and we recover 
the main result \eqref{lo result}; if $h \ll \mu^{1/2}$, $\Dtip$ is negligible and we recover 
the shallow-groove result \eqref{slip shallow}.

In what follows, we proceed to formalize the intuitive calculation leading to \eqref{dist intuitive slip}, making use of the distinguished limit
\begin{equation}\label{distinguished limit}
\mu,h\to0 \quad \text{with} \quad H=h/\mu^{1/2} \text{ fixed}.
\end{equation}
This is followed by the calculation of the next-order correction to obtain a two-term approximation to the slip length, analogous to \eqref{lam 2-term result}.

\subsection{Asymptotic regions and problem formulation}

Since the depth $h$ is $\ord(\mu^{1/2})$, and the two terms in Laplace's equation \eqref{Laplace both} balance when $x$ and $y$ are of comparable size, we anticipate the emergence of an intermediate asymptotic region, of $\ord(\mu^{1/2})$ extent, near the ridge. This is also hinted at by the results shown in Fig.~\ref{fig:numerical}(b): In the shallow interior domain, the variation is predominantly in the $y$-direction (as seen in the form of nearly horizontal contours), except within a distance of order $h$ about the ridge, representing the intermediate region.

We thus introduce the stretched intermediate coordinates
\begin{equation}
    (X,Y) = \frac{(x,y)}{\mu^{1/2}},
\end{equation}
for which the bottom boundary $y = -h $ becomes $Y = -H$. We continue to use $w^+(x,y)$ to denote the velocity in the exterior outer domain. Since the interior outer domain is shallow, we use (with a slight abuse of notation) $w^-(x,Y)$ to denote the velocity field there. For the intermediate region, we introduce the notation $W^\pm(X,Y)$, and for the exponentially small tip regions, we continue to use $\omega^\pm(\sigma,\phi)$ [where $\sigma$ is defined in \eqref{varrho def}]. 

On the outer and intermediate scales, the shear-stress continuity condition  \eqref{continuous stress} becomes
\begin{subequations}\label{dist continuous stress}
\refstepcounter{equation}\label{dist continuous stress outer}
\refstepcounter{equation}\label{dist continuous stress intermediate}
\begin{equation}\tag{\ref*{dist continuous stress}a,b}
    \pd{w^+}{y} = \mu^{1/2} \pd{w^-}{Y} \quad \text{at} \quad y=0, \qquad  \pd{W^+}{Y} = \mu \pd{W^-}{Y} \quad \text{at} \quad Y=0.
\end{equation}
\end{subequations}
For the total drag integral \eqref{drag both}, the integration interval $0 \leq x \leq 1$ now needs to be split [cf.~\eqref{balance split}] at two points $\delta_1$ and $\delta_2$, satisfying
\refstepcounter{equation}\label{delta3 range}
\begin{equation}\tag{\ref*{delta3 range}a,b,c}
    \delta_1 \ll \mu^{1/2}, \qquad \ln \frac{1}{\delta_1} \ll \mu^{-1/2},  \qquad \mu^{1/2} \ll \delta_2 \ll 1.
\end{equation}
\newcommand{\Dintermediate}{\Dtotal_{\mathrm{intermediate}}}
We therefore define $\Dtotal=\Dtip+\Dintermediate+\Douter$, where the respective contributions from the tip region ($0 < x < \delta_1$), the intermediate region ($\delta_1 < x < \delta_2$), and the outer region ($\delta_2 < x < 1$), are then given [after changes of variables, analogous to those  in \eqref{lo tip contribution}] by
\begin{subequations}\label{balance split3}
\begin{align}
\label{balance split3 tip}
\Dtip &= \mu^{1/2} \int_{-\infty}^{-\mu^{1/2}\ln(1/\delta_1)} \left.\pd{\omega^-}{\phi}\right|_{\phi=0}\,\intd{\sigma},
\\ 
\label{balance split3 intermediate}
\Dintermediate &= \mu \int_{\delta_1/\mu^{1/2}}^{\delta_2/\mu^{1/2}} \left.\pd{W^-}{Y}\right|_{Y=0}\,\intd{X},
\\
\label{balance split3 outer}
\Douter &= \mu^{1/2} \int_{\delta_2}^1 \left.\pd{w^-}{Y}\right|_{Y=0}\,\intd{x}.
\end{align}
The force balance \eqref{balance int} is then 
\begin{equation}
\label{balance split3 total}
\Dtip + \Dintermediate + \Douter = 1.
\end{equation}
\end{subequations}
Since the velocity fields are $\ord(\mu^{-1/2})$, we see that it is the tip and outer contributions that now contribute at leading order. The outer contribution, which is new at that order, is due to the shallowness of the groove, which has increased the shear stress 
from $\ord(\mu^{1/2})$ to $\ord(1)$. [The shear stress is also $\ord(1)$ in the intermediate region, but the horizontal extent of that region is only $\ord(\mu^{1/2})$.] 

We now proceed with the analysis, roughly following  the steps of Sec.~\ref{sec:Nearly inviscid}.

\subsection{Leading-order slip length}

As in expansions \eqref{outer expansion w} and \eqref{tip expansions}, we expand the velocity fields $w^\pm$, $W^\pm$ and $\omega^\pm$ in half-powers of $\mu$. 

Starting again with the leading-order outer exterior solution $w_{-1}^+$, continuity of shear stress \eqref{dist continuous stress outer} yields 
$\partial w_{-1}^+/\partial y = 0$ at $y=0$, 
which is the same as before [i.e., \eqref{outer stress -1}]. Since the other equations governing $w_{-1}^+$ also remain the same as in Sec.~\ref{sec:uniform ext},  the uniform solution \eqref{uniform ext} remains valid. 

Next, on the intermediate scale, the leading-order exterior field $W_{-1}^+$ satisfies Laplace's equation [cf.~\eqref{Laplace both}] and symmetry at $X=0$ [cf.~\eqref{symmetry+ x=0}]. Continuity of shear stress \eqref{dist continuous stress intermediate} yields $\partial W_{-1}^+/\partial Y = 0$ at $Y=0$, 
while matching to the outer solution $w_{-1}$ requires
$W_{-1}^+ \sim \lambda_{-1}$ as $X^2+Y^2 \to \infty$. Hence, the solution is simply an extension of \eqref{uniform ext},
\begin{equation}\label{uniform lo velocity intermediate}
W^+_{-1} \equiv \lambda_{-1}. 
\end{equation}

With the intermediate approximation \eqref{uniform lo velocity intermediate} having the same form as the outer approximation \eqref{uniform ext}, 
the leading-order inner tip approximation satisfies the same conditions as in Sec.~\ref{sec:tip solution}, including the matching condition \eqref{trivial match}.
Hence, the tip solution \eqref{omega lo} remains unchanged, as does its contribution \eqref{lo tip contribution result} to the total drag force \eqref{balance split3}, i.e., $\Dtip \sim \lambda_{-1}$.

The contribution $\Douter$ requires calculating the leading-order outer interior approximation $w_{-1}^-$. Laplace's equation \eqref{Laplace both} at leading order gives
\begin{equation}\label{dist w-1- ode}
    \pd{^2w_{-1}^-}{Y^2}=0,
\end{equation}
while no slip on the bottom [cf.~\eqref{noslip bottom}] and continuity with the exterior approximation \eqref{uniform ext} [cf.~\eqref{continuous velocity}] provide the boundary conditions 
\refstepcounter{equation}\label{dist w-1- bcs}
\begin{equation}\tag{\ref*{dist w-1- bcs}a,b}
    w_{-1}^- = 0 \quad \text{at} \quad Y=-H
, \qquad
    w_{-1}^- = \lambda_{-1} \quad \text{at} \quad Y=0.
\end{equation}
The solution is thus analogous to the shallow-groove approximation \eqref{1D profiles -},
\begin{equation}\label{dist w-1- solution}
    w_{-1}^- = \lambda_{-1}\left(1 + \frac{Y}{H}\right).
\end{equation}
[This solution trivially satisfies the symmetry condition at $x=1$. It does not have to satisfy the no-slip condition \eqref{noslip ridge} due to the presence of the intermediate region.] Hence, the contribution \eqref{balance split outer} to the drag from the outer region is 
\begin{equation}\label{D outer H}
    \Douter \sim \frac{\lambda_{-1}}{H}.
\end{equation}

Substituting the two $\ord(1)$ contributions \eqref{lo tip contribution result} and \eqref{D outer H} into \eqref{balance split3 total} 
and solving for $\lambda_{-1}$ yields the result
\begin{equation}\label{lam-1 dist}
    \lambda_{-1} = \frac{H}{1+H},
\end{equation}
from which we conclude that
\begin{equation}
\label{lam dist 1-term}
\lambda\sim \mu^{-1/2}\frac{H}{1+H} \quad \text{for } \mu\ll1 \text{ with fixed } H=h/\mu^{1/2},
\end{equation}
confirming the intuitively obtained leading-order result \eqref{dist intuitive slip}.

\begin{figure}[t!]
\begin{center}
\includegraphics[width=0.95\textwidth]{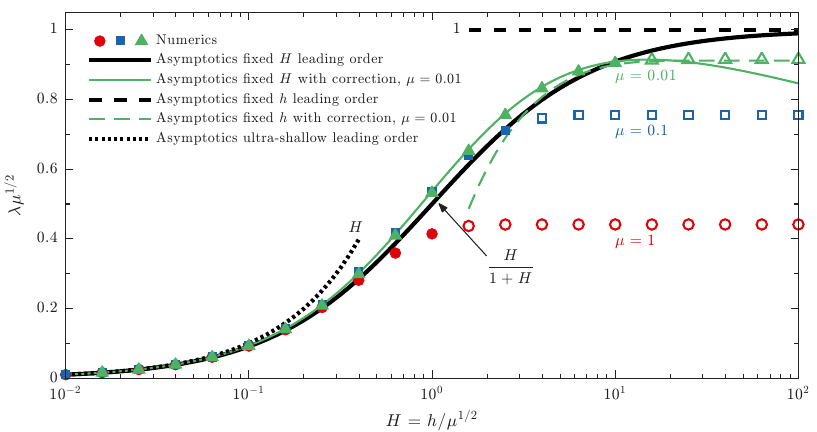}
\caption{Variation of the slip length (scaled as $\lambda \mu^{1/2}$) with scaled groove depth $H = h/\mu^{1/2}$ for various values of the viscosity ratio $\mu$. Symbols show numerical results, with open   symbols indicating $h>1$. The thick solid curve portrays the leading-order result \eqref{lam dist 1-term} in the distinguished limit $\mu,h \ll 1$ with fixed $H$. The thick dashed curve shows the leading-order result \eqref{lo result} in the original limit $\mu \ll 1$ with fixed $h$. The corresponding thin curves show the two-term approximations \eqref{lam dist 2-term} and \eqref{lam 2-term result}, evaluated for $\mu = 0.01$. The thick dotted curve shows the leading-order result \eqref{slip shallow} in the ultra-shallow-groove limit $h \ll 1$ with fixed $\mu$. (Under the rescalings of the axes, the leading-order approximations are independent of $\mu$.) 
%The two-term approximations \eqref{lam 2-term result}, for fixed $h$, and \eqref{lam dist 2-term}, for fixed $H$, are shown by the thin dashed and solid curves, respectively.   
%for fixed $h$ [cf.~\eqref{lam 2-term result} ] and fixed $H$  Thin solid and dashed curves show the two-term approximations \eqref{lam dist 2-term} and \eqref{lam 2-term result} 
%The corresponding thin curves show the two-term approximations \eqref{lam dist 2-term} and \eqref{lam 2-term result}, evaluated for $\mu = 0.01$.
} 
\label{fig:lambda_dist}
\end{center}
\end{figure}

In Fig.~\ref{fig:lambda_dist} we show the variation of the scaled slip length $\lambda \mu^{1/2}$ with the scaled groove depth $H$. The thick solid curve shows the leading-order approximation \eqref{lam-1 dist}, while the symbols depict numerical solutions for various values of $\mu$, obtained from our numerical scheme. The two agree well, provided that $h < 1$. In plotting numerical results, we employ filled  symbols for $h<1$ and open symbols for $h>1$; in the latter case, we do not expect approximation \eqref{lam dist 1-term} to hold. For perspective, we also depict (thick dashed curve) the original leading-order approximation \eqref{lo result} and (thick dotted curve) the ultra-shallow-groove approximation \eqref{slip shallow}, which are recovered by forming the respective limits $H \gg 1$ or $H \ll 1$ of \eqref{lam-1 dist}.

\subsection{Beyond leading-order slip length}
Continuing with the distinguished limit \eqref{distinguished limit}, 
in Appendix~\ref{app:dist0} we go beyond the preceding leading-order analysis, obtaining the two-term approximation
\begin{equation} \label{lam dist 2-term}
\lambda \sim \mu^{-1/2}\frac{H}{1+H} +\frac{2}{\pi}\left(\frac{H}{1+H}\right)^2 \ln\frac{1}{\mu^{1/2}H}\quad \text{for } \mu\ll1 \text{ with fixed } H = h/\mu^{1/2}.
\end{equation}
We note that the second  term includes contributions at both $\mathrm{ord}(\ln \mu)$ and $\mathrm{ord}(1)$; the asymptotic error in \eqref{lam dist 2-term} is \emph{algebraically} small. 

The two-term approximation \eqref{lam dist 2-term} is shown as a thin solid curve in Fig.~\ref{fig:lambda_dist} for $\mu=0.01$. As expected, it provides a better approximation than the one-term distinguished approximation (thick solid curve). For the same viscosity value, we also show (thin dashed curve) the 
two-term approximation \eqref{lam 2-term result}, corresponding to the original limit $\mu \ll 1$ with $h$ fixed. It is evident that the combination of \eqref{lam 2-term result} and \eqref{lam dist 2-term} practically describes the entire range of $h$.

In the limit $H\ll1$ the result \eqref{lam dist 2-term} simplifies to
\begin{equation}
\lambda \sim \frac{h}{\mu} - \frac{h^2}{\mu^{3/2}},
\end{equation}
which constitutes an extension of the shallow-groove result \eqref{slip shallow}. In the limit $H\gg1$ we get
\begin{equation}
\lambda \sim \mu^{-1/2} - \frac{1}{h} + \frac{2}{\pi}\ln\frac{1}{h},
\end{equation}
in agreement with the small-$h$ expansion \eqref{lam0 small h} of the original result \eqref{lam 2-term result}.

\section{Concluding remarks} \label{sec:conclude}
We have investigated the canonical problem of longitudinal shear over a liquid-infused grating formed of infinitely thin ridges. With lengths normalized by the grating half-period, the slip length $\lambda$ depends upon two dimensionless parameters: the interior-to-exterior viscosity ratio $\mu$ and the groove depth $h$. 
This assumes that the ridges are infinitely thin, that the liquid-liquid interfaces are flat and pinned to the ridge tips, and that the flow is laminar. (Since the flow is unidirectional, the inertial terms trivially vanish for any Reynolds number \citep{Batchelor:book}.) Our analysis has focused on the nearly-inviscid limit $\mu\ll1$, which is singular in the above zero-solid-fraction scenario. In that limit, we have shown that $\lambda$ possesses the asymptotic expansion 
\begin{equation}\label{conc exp m}
\lambda \sim \mu^{-1/2} + \lambda_0(h) \quad \text{as} \quad \mu\searrow0,
\end{equation}
where the $\ord{(1)}$ correction $\lambda_0(h)$ is provided by \eqref{lam0 arb h} in terms of an elliptic integral. Its small-$h$ behavior \eqref{lam0 small h} reveals a breakdown in the asymptotic expansion when $h$ becomes comparable to $\mu^{1/2}$. Proceeding with the distinguished shallow-grooves limit of $\mu\ll1$ and $H=h/\mu^{1/2}$ fixed, we have obtained the closed-form two-term approximation \eqref{lam dist 2-term}, which again starts at $\ord(\mu^{-1/2})$ but now with the leading term depending upon $H$. For $H\ll1$, namely, for extremely shallow grooves, this leading term reduces to $h/\mu$, revealing a transition of the scaling of $\lambda$ with $\mu$ across the distinguished regime. We have found excellent agreement between our asymptotic approximations and numerical solutions based upon a boundary-integral formulation.

As discussed in the Introduction, the $\mu^{-1/2}$ scaling of the effective slip length is at odds with both intuitive scaling arguments and previous heuristic models; these predict the scaling $\mu^{-1}$, which, as noted above, is only valid for extremely shallow grooves. Our analysis reveals that the $\mu^{-1/2}$ scaling arises from a unique asymptotic structure where the velocity of both liquids varies rapidly in exponentially small regions about the tips of the infinitely thin ridges. In particular, these regions resolve the incompatibility between a uniform stream in the exterior domain --- associated with an asymptotically large effective slip length and a nearly inviscid lubricant --- and the need to satisfy a no-slip condition at the tips. For $\mu\ll1$, the net viscous force supported by the liquid-liquid interface, which balances the imposed far-field shear, is dominated by the contribution from these exponentially small regions. This contradicts the naive scaling argument mentioned in the Introduction, where the viscous force was presumed to be associated with a geometric length scale, independent of $\mu$. As the groove depth $h$ is decreased, however, the contribution to the net viscous force from the groove  scale becomes increasingly important. In fact, the distinguished limit where $H=h/\mu^{1/2}$ is held fixed corresponds to the regime where the contributions from the exponential tip regions and the groove scales become comparable. Only for extremely shallow grooves, $H\ll1$, does the contribution from the groove scale become dominant and  the intuitive $\mu^{-1}$ scaling for the effective slip length is recovered. 

Admittedly, the leading-order role played by exponentially small regions near the ridge tips suggests that the theory developed herein for infinitely thin ridges and a nearly inviscid lubricant may be sensitive to perturbations. In particular, it is clear from our analysis that even an exceedingly small solid fraction $\epsilon$ --- exponentially small in $\mu$ --- may induce a leading-order effect by modifying the inner problem governing those regions. More specifically, we expect finite-solid-fraction effects to become important once the ridge thickness is comparable to the exponential length scale set by the small viscosity ratio, corresponding to the condition that $\mu^{1/2}\ln \epsilon $ is of order unity. In that distinguished limit, the order of magnitude of the effective slip length must still be $\mu^{-1/2}$. At larger (but still small) solid fractions, we expect a transition to the classical $\ln \epsilon $ scaling for an inviscid lubricant \citep{Philip:72}. Finally, for non-small solid fractions, the nearly inviscid limit ceases to be singular; indeed, that  scenario was studied by \citet{Crowdy:17:Perturbation} via a regular perturbation expansion. 

Another important perturbation involves our assumption that the liquid-liquid interface is pinned to the ridge tips: in some scenarios, the interface may instead lie slightly above or below them. Similar to the small-solid-fraction scenario, our analysis for infinitely thin ridges and a nearly inviscid lubricant hints to a leading-order effect once the vertical displacement of the interface becomes comparable to the exponentially small scale set by the lubricant viscosity. The case where the interface lies above the microstructure is primarily relevant to liquid-infused surfaces (SLIPS) in an ``encapsulated state,'' which is known to facilitate high drop mobility \citep{Smith:13}. In that case, the limit of an inviscid lubricant is singular even for a finite solid fraction, though it is intuitive that the scaling of the effective slip length would become more singular at sufficiently small solid fraction. The alternative case, where the exterior liquid partially invades the grooves, is relevant to superhydrophobic surfaces in the scenario where the liquid-gas menisci are depinned owing to a sufficiently high liquid pressure. In the latter scenario, the protrusion of the ridges into the liquid renders the inviscid-lubricant limit regular even for infinitely thin ridges \citep{Crowdy:21}; the pertinent singular limit is then that of a small invasion depth \citep{Yariv:23:tr,Yariv:23:highly,Yariv:24:grating}. 

Beyond addressing the singular perturbations discussed above, it would be valuable to extend the present theory to other microstructure geometries and flow configurations. With regards to the grating geometry, preliminary analysis suggests that the asymptotic structure of the transverse problem is similar to that found here for the longitudinal problem. However, a detailed analysis would entail solving more complex variants of the reduced Stokes-flow problems associated with this asymptotic structure (e.g., the local analysis near the tip ridges, and the driven-cavity problem) --- now governed by the biharmonic equation instead of Laplace's equation. Last, we comment on another common microstructure geometry: a doubly periodic array of posts protruding from a flat solid substrate. In contrast to the grating geometry, a finite lubricant viscosity does \emph{not} regularize here the effective slip length problem at zero solid fraction (i.e., infinitely thin posts). Accordingly, it would not make sense to attempt the limit process considered in this paper, of small lubricant viscosity at zero solid fraction. Instead, it would be of interest to study the asymptotic regime where both the solid fraction and the lubricant viscosity are small. This would constitute a generalization of  preceding small-solid-fraction analyses of the posts configuration which were limited to an inviscid lubricant \citep{Davis:10,Schnitzer:18:Fakir}.

\appendix

\section{Numerical method}\label{app:numerical}

Our numerical method is based on expressing the velocity and shear stress at the interface in terms of an auxiliary field $q(x)$ defined on $0<x<1$ using boundary integrals. These relations may be expressed as linear maps $M$, $M^+$ and $M^-$ (which are represented by  matrices after discretization) satisfying 
\begin{subequations}\label{num matrix def}
\refstepcounter{equation}\label{num matrix def M}
\refstepcounter{equation}\label{num matrix def M-}
\refstepcounter{equation}\label{num matrix def M+}
\begin{equation}\tag{\ref*{num matrix def}a,b,c}
	w^\pm(x,0) = M[q(x)], \quad \pd{w^-}{y}(x,0) = M^-[q(x)], \quad \pd{w^+}{y}(x,0) = M^+[q(x)] + 1.
\end{equation}
\end{subequations}
For any given value of $h$, we compute these matrices. Then, we calculate $q(x)$ for any desired value of $\mu$ by solving the linear equation
\begin{gather}\label{num equation}
	M^+[q(x)] + 1 = \mu\,M^-[q(x)],
\end{gather}
representing continuity of stress \eqref{continuous stress}, using MATLAB's built-in matrix solver. We then evaluate $w^\pm(x,0)$ using \eqref{num matrix def M}, and finally obtain the slip length from relation \eqref{slip integral}.

\subsection{Derivation of the boundary integrals}\label{app:numerical bi}

We first consider the interior solution, $y < 0$, for the case of infinitely deep grooves, $h = \infty$, and seek the field generated by a point source of strength $-2$ located on the interface at $(x,y) = (x',0)$ where $0 < x' < 1$. In the absence of lateral boundary conditions, a solution can be written as [cf.~\eqref{w0tilde+ source} below]
\begin{gather}\label{bi source}
    G_0(x-x',y) = -\frac{1}{2\pi}\ln[(x-x')^2+y^2] + C,
\end{gather}
where $C$ is an arbitrary constant. We use the method of images to construct a solution that satisfies the no-slip condition \eqref{noslip ridge} at $x=0$ and symmetry condition \eqref{symmetry -} at $x=1$, by adding image sources in $x<0$ and $x>1$ such that the resulting source distribution is anti-symmetric about $x=0$ and symmetric about $x=1$. 

Specifically, we add sources at $x = x' + 4n$ and $x = 2-x' + 4n$ and subtract sources at $x = -x' + 4n$ and $x = -2+x' + 4n$, where $n$ ranges over all integers. Each of these four sets forms an array with equal spacing $4$, for which the result [obtained by summing \eqref{bi source} with judicious choices of $C$ for each point, analogously to \eqref{w0tilde+ series 4}] is
\begin{gather}\label{bi G4}
	G_4(x,y;x'') = -\frac{1}{2\pi}\ln\left[\sin^2 \frac{\pi (x-x'')}{4} + \sinh^2 \frac{\pi y}{4}\right],
\end{gather}
where $x''$ takes the values $x'$, $2-x'$, $-x'$ and $-2-x'$, respectively, and a suitable value of the arbitrary additive constant has been chosen. The solution is then given by the superposition
\begin{gather}\label{bi G}
  G(x,y;x') = G_4(x,y;x') + G_4(x,y;2-x') - G_4(x,y;-x') - G_4(x,y;-2+x').
\end{gather}
[Note that changing the additive constant in \eqref{bi G4} does not alter this result.]

We now let $q(x')$ represent the density of point sources in $0 < x' < 1$, and construct the solution
\begin{gather}\label{bi w-}
	w^-(x,y) = \int_0^1 G(x,y;x')q(x')\,\intd{x'},
\end{gather}
which satisfies \eqref{Laplace both}, \eqref{symmetry -}, \eqref{noslip ridge} and \eqref{decay interior}. Thus, for $h=\infty$, the interfacial quantities can be expressed as
\begin{subequations}\label{bi w- deep interface}
\refstepcounter{equation}\label{bi w- deep interface velocity}
\refstepcounter{equation}\label{bi w- deep interface shear}
\begin{equation}\tag{\ref*{bi w- deep interface}a,b}
	w^-(x,0) = \int_0^1 G(x,0;x')q(x')\,\intd{x'}, \qquad \pd{w^-}{y}(x,0^-) = q(x),
\end{equation}
where
\begin{equation}
	G(x,0;x') = \frac{1}{2\pi}\ln \frac{\sin^2 \frac{\pi(x+x')}{4} \sin^2 \frac{\pi(x+2-x')}{4}}{\sin^2 \frac{\pi(x-x')}{4} \sin^2 \frac{\pi(x+x'-2)}{4}}.
\end{equation}
\end{subequations}

Considering now the case of finite $h$, we observe that the no-slip boundary condition \eqref{noslip bottom} at $y=-h$ can be satisfied by simply introducing an opposite image distribution at $y=-2h$, so we can construct the solution
\begin{gather}
	w^-(x,y) = \int_0^1 \left[G(x,y;x') - G(x,-2h-y;x')\right]q(x')\,\intd{x'},
\end{gather}
to Eqs.~\eqref{Laplace both}, \eqref{symmetry -} and \eqref{noslip both}. This yields the interfacial quantities
\newcommand{\sss}{g}
\begin{subequations}\label{bi w- arb h interface}
\begin{align}
    \label{bi w- arb h interface velocity}
	w^-(x,0) &= \int_0^1 \left[G(x,0;x') - G(x,-2h;x')\right]q(x')\,\intd{x'},
	\\
    \label{bi w- arb h interface shear}
	\pd{w^-}{y}(x,0) &= q(x) + \int_0^1 \pd{G}{y}(x,-2h;x')q(x')\,\intd{x'},
\end{align}
where 
\begin{align}
    G(x,-2h;x') &= \frac{1}{2\pi} \ln \frac{\sss(x+x')\,\sss(x+2-x')}{\sss(x-x')\,\sss(x+x'-2)},
    \\
    \pd{G}{y}(x,-2h;x') &= \frac{\coth(\pi h/2)}{4} \left[\frac{1}{\sss(x-x')} + \frac{1}{\sss(x+x'-2)} \right. 
    \notag \\ & \hspace{7em} \left. {}-\frac{1}{\sss(x+x')}-\frac{1}{\sss(x+2-x')}\right],
\end{align}
\end{subequations}
in which $\sss(x'') = 1 + [\sin^2(\pi x''/4)/\sinh^2(\pi h/2)]$.  
[For $h=\infty$, we have $\sss(x'') = 1$ so that \eqref{bi w- arb h interface} reduce to \eqref{bi w- deep interface}. Hence, we no longer have to treat the case $h=\infty$ separately.]

Next, we consider the exterior solution, with $y > 0$. We subtract a linear profile from $w^+$ to obtain a field
\begin{equation}
    \bar{w}^+(x,y) = w^+(x,y) - y,
\end{equation}
which satisfies the same equations as $w^+$, but remains bounded as $y \to \infty$ [instead of \eqref{far refined}]. The interfacial shear stress, to be used in \eqref{continuous stress}, becomes
\begin{equation}
    \pd{w^+}{y}(x,0) = \pd{\bar{w}^+}{y}(x,0) + 1.
\end{equation}

We treat the value of $\bar{w}^+(x,y)$ on the interface $y=0$ as given, namely by $w^-(x,0)$ [due to continuity \eqref{continuous velocity}]. Since $\bar{w}^+(x,y)$ satisfies the symmetry conditions \eqref{symmetry+ both} at $x=0$ and $x=1$, the solution should involve a superposition of two arrays of point sources (of strength $-2$, say) with period $2$, which analogously to \eqref{bi G4} are given by
\begin{equation}
    G_2(x,y;x'') = -\frac{1}{2\pi}\ln\left[\sin^2 \frac{\pi(x-x'')}{2} + \sinh^2 \frac{\pi y}{2}\right],
\end{equation}
with $x''$ taking the values $x'$ and $-x'$. In light of \eqref{bi w- deep interface shear}, the construction 
\begin{equation}\label{bi wbar+}
    \bar{w}^+(x,y) = -\pd{}{y} \int_0^1 \left[G_2(x,y;x') + G_2(x,y;-x')\right]w^-(x',0)\,\intd{x'}
\end{equation}
similarly satisfies $\bar{w}^+(x,0^+) = w^-(x,0)$, as desired. Further, it satisfies Eqs.~\eqref{Laplace both} and \eqref{symmetry+ both}, and remains bounded as $y \to \infty$, so it is the required solution.

In order to evaluate the interfacial shear of \eqref{bi wbar+}, we use the fact that the integral in \eqref{bi wbar+} is harmonic, to write
\begin{gather}
    \pd{\bar{w}^+}{y}(x,y) = \pd{^2}{x^2} \int_0^1 \left[G_2(x,y; x') + G_2(x,y;-x')\right]w^-(x',0)\,\intd{x'}.
\end{gather}
Although this could be evaluated directly at $y=0$, we found that the numerical results become more accurate if we move one derivative inside the integral before taking the limit $y \searrow 0$. This yields
\begin{gather}\label{bi w+ interface shear}
    \pd{\bar{w}^+}{y}(x,0) = -\frac{1}{2} \pd{}{x} \int_0^1 \left[\cot\frac{\pi(x-x')}{2} + \cot\frac{\pi(x+x')}{2}\right]w(x',0)\,\intd{x'},
\end{gather}
where the Cauchy principal value is taken at the singularity $x' = x$. 

Thus, the linear maps defined in \eqref{num matrix def} are obtained as follows: \eqref{bi w- arb h interface velocity} and \eqref{bi w- arb h interface shear} provide $M$ and $M^-$, respectively, while \eqref{bi w+ interface shear} combined with \eqref{bi w- arb h interface shear} provides $M^+$. It remains to detail how the system is discretized to become a matrix equation.

\subsection{Discretization}

\newcommand{\xmin}{x_{\mathrm{min}}}
In order to solve the equation \eqref{num equation} numerically, we need to discretize the domain $0 \leq x \leq 1$. There is no dependence on $y$, since the equations have been mapped onto the interface $y=0$, with $q(x)$ being the unknown to solve for. 

We use an exponentially stretched grid to resolve the small scale near the ridge tip $x=0$. This grid is chosen to have a ``relative'' resolution of $\Delta = 0.002$ down to a minimum of $\xmin = 10^{-10}$, such that $\ln(x+\xmin)$ ranges from $\ln(\xmin)$ to $\ln(1+\xmin)$ in approximately uniform steps of $\Delta$.

Having divided the interval $0 < x < 1$ into cells, we sample $q(x)$, $w^\pm(x,0)$ and $\partial w^\pm/\partial y(x,0)$ at the cell midpoints. The integrals in \eqref{bi w- arb h interface} can be directly evaluated using the midpoint rule. However, in order to improve the accuracy of \eqref{bi w- arb h interface velocity}, we remove the logarithmic singularity in $G(x,0;x')$ at $x'=x$ using the integral
\begin{equation}\label{bi singularity removal}
\int_0^1 \frac{1}{2\pi}\ln\left[\left(\frac{4}{\pi(x-x')}\right)^2\right]\,q(x)\,\intd{x'} = \frac{1}{\pi}\left[1 + \ln \frac{4}{\pi} - x \ln x - (1-x)\ln(1-x)\right]q(x).
\end{equation}
By subtracting the integrand in \eqref{bi singularity removal} from the integrand in \eqref{bi w- arb h interface velocity}, and then adding back the right-hand side of \eqref{bi singularity removal}, we obtain an expression for $w^-(x,0)$ in terms of $q(x)$ where the integrand is bounded, resulting in second-order accuracy when evaluated using the midpoint rule. In order to further increase accuracy for $x$ close to $0$ or $1$ where the logarithmic singularity at $x'=-x$ or $x'=2-x$ is close to (albeit still outside of) the range of integration, we also perform the corresponding removal procedure for those two singularities (regardless of the value of $x$).

The derivative in \eqref{bi w+ interface shear} is evaluated using central finite differences. This requires the integral to be evaluated with $x$ at the cell boundaries. We continue to evaluate the integral using the midpoint rule, i.e., with $x'$ at the cell midpoints. Due to $x$ and $x'$ being staggered in this way, direct evaluation of the integral results in the Cauchy principal value being obtained automatically for the singularity at $x'=x$. However, we improve the accuracy of the result by subtracting the identity
\begin{gather}\label{bi singularity removal 2}
    \int_0^1 \left[\cot\frac{\pi(x-x')}{2} + \cot\frac{\pi(x+x')}{2}\right]w(x,0) \,\intd{x'} = 0.
\end{gather}
This results in an integral of unchanged value but now finite integrand, whose numerical evaluation is again accurate to second order.

As a result of \eqref{bi singularity removal}, \eqref{bi singularity removal 2}, and the use of the midpoint method and central differencing, the resulting discrete system is accurate to second order. We have verified that this is the case, by varying $\Delta$ between $0.002$ and $0.1$ and checking that the error (compared with the result for the smallest value of $\Delta$) scales like $\Delta^2$.

For small $\mu$, even exponential stretching is unable to resolve down to the inner scale where $\ln x = \ord(\mu^{-1/2})$. Hence, we make use of the local behavior \eqref{local interface}, and prescribe the values of $q(x)$ for $x < \sqrt{\xmin} = 10^{-5}$ to satisfy
\refstepcounter{equation}\label{num local}
\begin{equation}\tag{\ref*{num local}a,b}
    q(x) = q(\sqrt{\xmin}) \left(\frac{x}{\xmin}\right)^{\alpha-1}, \qquad w(x,0) = w(\sqrt{\xmin},0) \left(\frac{x}{\xmin}\right)^{\alpha},
\end{equation}
where $\alpha = (2/\pi)\arctan \mu^{1/2}$ is given by \eqref{local alpha}. Having verified that \eqref{num local} yields accurate results for moderate values of $\mu$ when compared with the original formulation without \eqref{num local}, we then make use of \eqref{num local} for all values of $\mu$ (even when not necessary, for simplicity).

\section{The correction to the outer approximation in the exterior domain} \label{app:w0tilde+}

We seek the harmonic function $\tilde{w}_0^+$ [defined in \eqref{w0 shift}] that satisfies the homogeneous Neumann conditions [cf.~\eqref{symmetry+ both} and \eqref{outer stress 0}]
\begin{subequations}\label{w0tilde+ app Neumann}
\begin{gather}\tag{\ref*{w0tilde+ app Neumann}a,b}
    \pd{\tilde{w}_0^+}{x} = 0 \quad \text{at} \quad x = 0,1, \qquad \pd{\tilde{w}_0^+}{y} = 0 \quad \text{at} \quad y=0;
\end{gather}
\end{subequations}
and the far-field and tip conditions [cf.~\eqref{refined far} and \eqref{source}],
\begin{subequations}\label{w0tilde+ app far tip}
\refstepcounter{equation}\label{w0tilde+ app far}
\refstepcounter{equation}\label{w0tilde+ app tip}
\begin{gather}\tag{\ref*{w0tilde+ app far tip}a,b}
    \tilde{w}_0^+ = y + o(1) \quad \text{as} \quad y \to \infty, \qquad \tilde{w}_0^+ \sim \frac{2}{\pi}\ln\rho \quad \text{as} \quad \rho = \sqrt{x^2+y^2} \to 0. 
\end{gather}
\end{subequations}
We provide derivations using two different methods, as they are useful for generalizing the result for use in the other Appendixes. Both methods use the complex variable $z = x + \ii y$, and yield the result
\begin{gather}\label{w0tilde+ app result}
	\tilde{w}_0^+ = \frac{1}{\pi}\ln\left|4\sin^2 \frac{\pi z}{2}\right| = \frac{1}{\pi}\ln\left[4\left(\sin^2\frac{\pi x}{2}+\sinh^2\frac{\pi y}{2}\right)\right],
\end{gather}
which clearly satisfies the required conditions \eqref{w0tilde+ app Neumann}--\eqref{w0tilde+ app far tip}.

\subsection{Solution via method of images}

We employ the fundamental solution of the 2D Laplace's equation with an arbitrary additive constant $C$, 
\begin{gather}\label{w0tilde+ source}
    G_0(x,y) = \frac{1}{\pi}\ln(x^2+ y^2) + C.
\end{gather}
Representing a ``source of $4$'', it matches \eqref{w0tilde+ app tip}. In addition, it satisfies the Neumann conditions \eqref{w0tilde+ app Neumann} at $x=0$ and at $y=0$, but not at $x=1$. 

To obtain a solution that also satisfies the symmetry condition at $x=1$, which together with the symmetry condition at $x=0$ implies a period of 2, we need to superpose a periodic array of sources \eqref{w0tilde+ source} translated in $x$ by $2n$, where $n$ ranges over all integers, with the constants $C$ chosen judiciously to avoid the series diverging. We thus posit the solution
\refstepcounter{equation}\label{w0tilde+ series 1}
\begin{gather}\tag{\ref*{w0tilde+ series 1}a,b}
    \tilde{w}_0^+(x,y) = \sum_{n=-\infty}^{\infty} \left[\frac{1}{\pi}\ln\left[(x+2n)^2+y^2\right] + C_n\right],
    \quad
    C_n = -\frac{1}{\pi}\ln\left[(2n)^2\right] \ \text{for} \ n\neq 0,
\end{gather}
leaving $C_0$ undetermined so far. Introducing the complex variable $z = x + \ii y$, we rewrite \eqref{w0tilde+ series 1} as
\begin{gather}\label{w0tilde+ series 2}
    \tilde{w}_0^+(x,y) = C_0 + \frac{1}{\pi}\ln\left\{\left[\prod_{n=-\infty}^{-1} \frac{|z+2n|^2}{2n^2}\right]|z|^2 \left[\prod_{n=1}^\infty \frac{|z+2n|^2}{2n^2}\right]\right\}.
\end{gather}
We  pair up corresponding factors in each product to obtain
\begin{equation}\label{w0tilde+ series 3}
    \tilde{w}_0^+(x,y) = C_0 + \frac{1}{\pi} \ln \left|z \prod_{n=1}^\infty \frac{z^2 - 4n^2}{4n^2}\right|^2.
\end{equation}
Using the Weierstrass factorization formula for the sine function, this can be rewritten as
\begin{equation}\label{w0tilde+ series 4}
   \tilde{w}_0^+(x,y) = C_0 + \frac{1}{\pi}\ln \left|\frac{2}{\pi} \sin \frac{\pi z}{2}\right|^2.
\end{equation}
Reverting to $(x,y)$ and choosing $C_0$ to satisfy the far-field condition \eqref{w0tilde+ app far} then yields the result \eqref{w0tilde+ app result}.

\subsection{Solution via conformal mapping}\label{app:w0tilde+ mapping}

The problem \eqref{w0tilde+ app Neumann}--\eqref{w0tilde+ app far tip}, if extended symmetrically to $y<0$, is reminiscent of the classical irrotational-flow problem of narrow-slit influx into a 2D channel, which is solved using conformal mapping in the standard textbooks \cite{Brown:book}. We use a different mapping here, which generalizes more readily to the other problems solved in the next section.

\begin{figure}[hbtp]
\begin{center}
\includegraphics{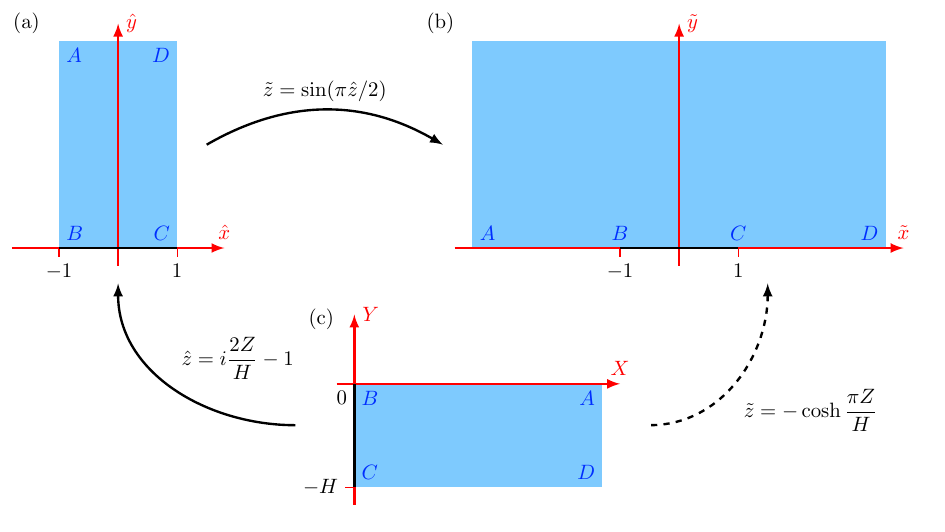}
\caption{Conformal maps from a semi-infinite strip (a,c) to the upper half plane (b). In Appendix~\ref{app:w0tilde+ mapping}, (a) is mapped to (b), with $\tilde{w}_0^+$ satisfying homogeneous Neumann conditions on all boundaries (except at the origin). In Appendix~\ref{app:w-1- deep}, (a) is mapped to (b), with Dirichlet conditions $w_{-1}^- = 0$ on $AB$ and $CD$, and $w_{-1}^-=1$ on $BC$. In Appendix~\ref{app:dist0 intermediate}, (c) is mapped via (a) to (b), with Dirichlet conditions $W_{-1}^-=1$ on $AB$ and $W_{-1}^-=0$ on $BC$ and $CD$.}
\label{fig:mapping}
\end{center}
\end{figure}

We consider a symmetric extension about $x=0$ to the domain $-1 < x < 1$ (with $y > 0$), resulting in the homogeneous Neumann conditions \eqref{w0tilde+ app Neumann} applying at the new boundaries $x=-1$, $x=1$ and $y=0$ (except at the singularity at the origin), while conditions \eqref{w0tilde+ app far tip} are retained. We embed the harmonic $\tilde{w}_0^+$ in the complex potential
\begin{equation}\label{complex potential *}
\Phi = \tilde{w}_0^+ + \ii \chi, 
\end{equation}
which is an analytic function of $z=x+\ii y$, and find the appropriate solution $\Phi$ by mapping the domain $-1 < x < 1$ (with $y>0$) to the upper half plane using the transformation [see Fig.~\ref{fig:mapping}(a,b)]
\begin{subequations}\label{w0tilde+ sinemap}
\refstepcounter{equation}\label{w0tilde+ sinemap z}
\refstepcounter{equation}\label{w0tilde+ sinemap x}
\refstepcounter{equation}\label{w0tilde+ sinemap y}
\begin{equation}\tag{\ref*{w0tilde+ sinemap}a,b,c}
    \tilde{z} = \sin \frac{\pi z}{2}, \qquad \tilde{x} = \sin \frac{\pi x}{2} \cosh \frac{\pi y}{2}, \qquad \tilde{y} = \cos \frac{\pi x}{2} \sinh \frac{\pi y}{2}.
\end{equation}
\end{subequations}
The boundaries at $x=\pm 1$ and $y=0$ map to the new real axis $\tilde{y} = 0$. In terms of the new variables (with a slight abuse of notation), $\tilde{w}_0^+$ satisfies 
\begin{subequations}\label{w0tilde+ mapped}
\begin{equation}\label{w0tilde+ mapped Neumann}
    \pd{\tilde{w}_0^+}{\tilde{y}} = 0 \quad \text{at} \quad \tilde{y} = 0,
\end{equation}
except at the origin. Conditions \eqref{w0tilde+ app far tip} transform, using \eqref{w0tilde+ sinemap}, to
\refstepcounter{equation}\label{w0tilde+ mapped far}
\refstepcounter{equation}\label{w0tilde+ mapped tip}
\begin{equation}\tag{\ref*{w0tilde+ mapped}b,c}
    \tilde{w}_0^+ = \frac{2}{\pi}\ln\left(2\tilde{\rho}\right) + o(1) \quad \text{as} \quad \tilde{\rho} \to \infty,
    \qquad
    \tilde{w}_0^+ \sim \frac{2}{\pi}\ln \tilde{\rho}\quad \text{as} \quad \tilde{\rho} \to 0,
\end{equation}
\end{subequations}
where $\tilde{\rho} = \sqrt{\tilde{x}^2 + \tilde{y}^2} = |\tilde{z}|$. Since $\ln (2\tilde{\rho})$ is the real part of $\ln(2\tilde{z})$, which is analytic in the upper half plane, it is clear that a solution to the problem \eqref{w0tilde+ mapped} is given by
\begin{subequations}\label{w0tilde+ mapped solution}
\refstepcounter{equation}\label{w0tilde+ mapped solution Phi}
\refstepcounter{equation}\label{w0tilde+ mapped solution w0tilde+}
\begin{equation}\tag{\ref*{w0tilde+ mapped solution}a,b}
    \Phi = \frac{2}{\pi}\ln(2\tilde{z}), \qquad \tilde{w}_0^+ = \frac{2}{\pi}\ln(2\tilde{\rho}).
\end{equation}
\end{subequations}
Restoring $z$ using \eqref{w0tilde+ sinemap z} yields the desired result \eqref{w0tilde+ app result}.

\section{Leading-order outer approximation in the interior domain}\label{app:w-1-}

In order to calculate the leading-order outer interior flow $w_{-1}^-$ for Sec.~\ref{sec:interior}, we introduce a shifted horizontal coordinate and flipped vertical coordinate,
\refstepcounter{equation}\label{w-1- coords}
\begin{equation}\tag{\ref*{w-1- coords}a,b}
    \hat{x} = x-1, \qquad \hat{y} = -y,
\end{equation}
and consider a full groove $-1 < \hat{x} < 1$ (with $\hat{y} > 0$) instead of imposing the symmetry condition \eqref{symmetry+ x=1} at $\hat{x}=0$. Thus, with a slight abuse of notation, we seek a harmonic function $w_{-1}^-(\hat{x},\hat{y})$ satisfying the Dirichlet conditions [cf.~\eqref{noslip both} and \eqref{interfacial'}]
\refstepcounter{equation}\label{w-1- app Dirichlet}
\begin{gather}\tag{\ref*{w-1- app Dirichlet}a,b}
    w_{-1}^- = 0 \quad \text{at} \quad \hat{x} = \pm 1 \quad \text{and at} \quad \hat{y} = h,
    \qquad 
    w_{-1}^- = 1 \quad \text{at} \quad \hat{y} = 0.
\end{gather}
We use conformal mapping as in Appendix~\ref{app:w0tilde+ mapping}, and embed $w_{-1}^-$ in a complex potential
\begin{equation}\label{complex potential int}
\Phi = w^-_{-1} + \ii \chi, 
\end{equation}
a function of $\hat{z} = \hat{x} + \ii \hat{y}$.

\subsection{Infinitely deep grooves} \label{app:w-1- deep}

In the limit of infinitely deep grooves, $h = \infty$, the no-slip condition at $\hat{y}=h$ is replaced by a decay condition as $\hat{y}\to \infty$ [see \eqref{decay interior}]. The resulting problem appears in standard textbooks, e.g.~as a heat conduction problem in a semi-infinite slab whose end is held at an elevated temperature compared with its sides \citep{Brown:book}. The solution is described here briefly for completeness. 

We again map the semi-infinite strip to the upper half plane via the transformation $\tilde{z} = \sin(\pi \hat{z}/2)$ [analogous to \eqref{w0tilde+ sinemap}, illustrated in Fig.~\ref{fig:mapping}]. Since Dirichlet conditions are conformally invariant, Eqs.~\eqref{w-1- app Dirichlet} become
\refstepcounter{equation}\label{w-1- deep Dirichlet}
\begin{gather}\tag{\ref*{w-1- deep Dirichlet}a,b}
    w_{-1}^-(\tilde{x},0) = \left\{\begin{array}{lll} 0 & \text{ for } & |\tilde{x}| > 1, \\ 1 & \text{ for } & |\tilde{x}| < 1, \end{array}\right.
		\qquad 
		w_{-1}^- \to 0 \quad \text{as} \quad \tilde{\rho} \to \infty.
\end{gather}
Since the imaginary part of
$\log \tilde{z} = \ln \tilde{\rho} + \ii \arg \tilde{z}$
takes the values $0$ and $\pi$ on the two halves of the real axis, it is clear that a solution to \eqref{w-1- deep Dirichlet} can be constructed as    
\refstepcounter{equation}\label{w-1- deep solution}
\begin{gather}\tag{\ref*{w-1- deep solution}a,b}
    \Phi = \frac{1}{\ii \pi}\left[\log(\tilde{z}-1)-\log(\tilde{z}+1)\right], \qquad w_{-1}^- = \frac{1}{\pi}\arg \frac{\tilde{z}-1}{\tilde{z}+1}.
\end{gather}
Converting to real form and reverting to $\hat{x}$ and $\hat{y}$ using the analog of \eqref{w0tilde+ sinemap} then yields
\begin{subequations}\label{w-1- deep solution real}
\begin{align}
    w_{-1}^- &= \frac{1}{\pi}\arctan \frac{2\tilde{y}}{\tilde{x}^2+\tilde{y}^2-1}
    \\
    &= \frac{1}{\pi}\arctan\frac{2\cos\frac{\pi \hat{x}}{2}\sinh\frac{\pi \hat{y}}{2}}{\sin^2 \frac{\pi \hat{x}}{2} + \sinh^2 \frac{\pi \hat{y}}{2} - 1},
\end{align}
\end{subequations}
where the range of the arctangent is taken to be $0$ to $\pi$. We can then evaluate the interfacial shear for $-1 < \hat{x} < 1$ as
\begin{equation}\label{w-1- deep shear}
    -\left.\pd{w_{-1}^-}{\hat{y}}\right|_{\hat{y}=0} = \frac{1}{\cos(\pi \hat{x}/2)},
\end{equation}
from which \eqref{interior shear -1} follows.

\subsection{Arbitrary groove depth}\label{sec:w-1- arb h} 

\begin{figure}[hbtp]
\begin{center}
\includegraphics{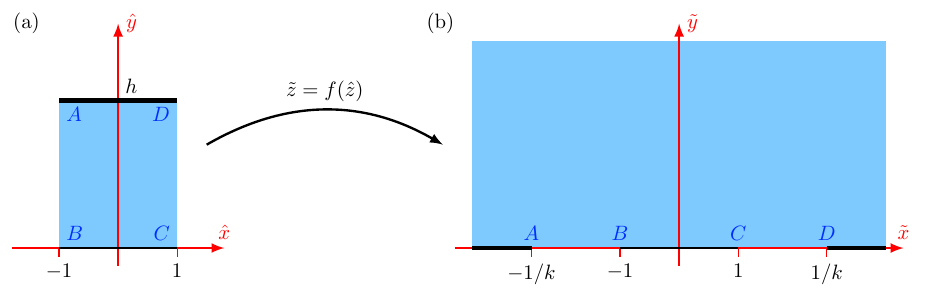}
\caption{Conformal map from a rectangle to the upper-half plane used in Appendix~\ref{sec:w-1- arb h}, whose inverse is a Schwarz--Christoffel map \eqref{arb h SC}. The boundary conditions are $w_{-1}^- = 0$ on all boundaries except $BC$, where $w_{-1}^-=1$ instead.}
\label{fig:mapping_sc}
\end{center}
\end{figure}

For an arbitrary groove depth $h$, we need the map $\tilde{z}=f(\hat{z})$ from the rectangular domain $-1 < \hat{x} <1$, $0 < \hat{y} < h$ to the upper half plane $\tilde{y} > 0$; see Fig.~\ref{fig:mapping_sc}. This is the inverse of a Schwarz--Christoffel map \citep[e.g.][]{Brown:book}, which due to symmetry in $\hat{x}$ and $\tilde{x}$ has the form
\begin{equation}\label{arb h SC}
    \hat{z} = f^{-1}(\tilde{z}) = -M\int_0^{\tilde{z}} \frac{\intd{\nu}}{(\nu+1/k)^{1/2}(\nu+1)^{1/2}(\nu-1)^{1/2}(\nu-1/k)^{1/2}}, 
\end{equation}
where the positive real constants $M$ and $k$ remain to be determined. We shall express $f^{-1}$ in terms of the incomplete and complete elliptic integrals of the first kind with parameter $m=k^2$, defined by
\begin{subequations}\label{arb h elliptic}
\refstepcounter{equation}\label{arb h elliptic incomplete}
\refstepcounter{equation}\label{arb h elliptic complete}
\begin{equation}\tag{\ref*{arb h elliptic}a,b}
    F(\tilde{z};m) = \int_0^{\tilde{z}} \frac{\intd{\nu}}{(1-\nu^2)^{1/2}(1-m\nu^2)^{1/2}}, \qquad K(m) = F(1;m).
\end{equation}
\end{subequations}
Comparison of \eqref{arb h SC} and \eqref{arb h elliptic incomplete} reveals that $f^{-1}(\tilde{z}) = kMF(\tilde{z};k^2)$. Next, the mapping of point $D$ in Fig.~\ref{fig:mapping_sc} yields the condition $1+\ii h = f(1/k)$, which after splitting the integral at $\nu=1$ and separating the real and imaginary parts becomes
\begin{equation}
    1 = kMK(k^2), \qquad h = kMK(1-k^2).
\end{equation}
Eliminating $M$, we deduce that $k$ depends on $h$ according to the equation
\begin{equation}
    \frac{K(1-k^2)}{K(k^2)} = h,
\end{equation}
and the (inverse) mapping \eqref{arb h SC} can be expressed as
\begin{equation}\label{arb h f-1}
    \hat{z} = f^{-1}(\tilde{z}) = \frac{F(\tilde{z};k^2)}{K(k^2)}.
\end{equation}
\newcommand{\sn}{\mathop{\mathrm{sn}}}%
The inverse of the function $F({}\cdot{};k^2)$ is the Jacobi elliptic sine function with parameter $m = k^2$, denoted $\sn({}\cdot{},k^2)$. Using this, we can express the forward mapping as
\begin{equation}\label{arb h SC forward}
    \tilde{z} = f(\hat{z}) = \sn(K(k^2)\hat{z};k^2).
\end{equation}

Next, we need to solve for $w_{-1}^-$ in the upper half $\tilde{z}$ plane. The interface on which $w_{-1}^-=1$ is mapped from $\hat{y}=0$ with $-1<\hat{x}<1$ to $\tilde{y}=0$ with $-1 < \tilde{x} < 1$, see Fig.~\ref{fig:mapping_sc}. The remaining portions of the rectangular boundary, on which $w_{-1}^-=0$, are mapped to the remainder of the real axis. Hence, in the $\tilde{z}$ plane, $w_{-1}^-$ again satisfies the Dirichlet conditions \eqref{w-1- deep Dirichlet}, and the  solution for $w_{-1}^-$ is thus given by \eqref{w-1- deep solution}.

The interfacial shear, however, is not given by \eqref{w-1- deep shear}, since the mapping \eqref{arb h SC forward} is different. From the chain rule, $\partial w_{-1}^-/\partial \hat{z} =(d f/d \hat{z})(\partial w_{-1}^-/\partial \tilde{z}) $, the shear is
\begin{equation}
    -\left.\pd{w_{-1}^-}{\hat{y}}\right|_{\hat{y}=0} = \frac{2}{\pi} \frac{1}{1 - f^2(\hat{x})}\frac{df(\hat{x})}{d\hat{x}}.
\end{equation}
Hence, the sought regularized drag integral \eqref{F def} becomes
\begin{subequations}
\begin{align}
    \mathcal{F} &= \frac{2}{\pi}\int_{-1}^0 \left[\frac{1}{1-f^2(\hat{x})}\frac{df(\hat{x})}{d\hat{x}} - \frac{1}{\hat{x}+1}\right] \,\intd{\hat{x}}
    \\ \label{arb h F step1}
    &= \frac{2}{\pi}\lim_{\hat{x}\to -1} \left[\frac{1}{2}\ln \frac{1-f(\hat{x})}{1+f(\hat{x})} + \ln(\hat{x}+1)\right].
\end{align}
\end{subequations}
To calculate the limit, we use \eqref{arb h f-1} and \eqref{arb h elliptic} to obtain 
\begin{equation}\label{arb h f-1 near -1}
    f^{-1}(-1 + \varepsilon) \sim -1 + \frac{\sqrt{2\varepsilon}}{K(k^2)\sqrt{1-k^2}} \quad \text{as} \quad \varepsilon \searrow 0,
\end{equation}
whereby 
\begin{equation}
    f(\hat{x}) \sim -1 + \frac{1-k^2}{2} K^2(k^2) (1+\hat{x})^2 \quad \text{as} \quad \hat{x} \searrow -1.
\end{equation}
Substitution into \eqref{arb h F step1} then yields the result \eqref{F arb h}.

\section{Higher-order approximation for the slip length in the distinguished shallow-groove limit}\label{app:dist0}

We seek to calculate the asymptotic correction to the leading-order result \eqref{lam-1 dist} in the distinguished limit for shallow grooves (Sec.~\ref{sec:dist}). 
We proceed to calculate the solution fields required to determine $\Dtip$, $\Douter$ and $\Dintermediate$ to $\ord(\mu^{1/2})$. Following \citet{Fraenkel:69}, we group together terms that are separated only by a logarithmic factor (in $\mu$), and ensure that all approximations involve discarding terms are algebraically small.

\subsection{Higher-order outer approximation in the exterior domain and inner approximation near the tip}\label{app:dist0 Stip}

We start by considering the $\ord(1)$ outer-scale problem in the exterior domain. Like in the calculation for fixed $h$ (see Sec.~\ref{sec:slip 0}), the field $w_0^+(x,y)$ satisfies Laplace's equation [cf.~\eqref{Laplace both}], homogeneous Neumann conditions at $x=0$ and $x=1$ [cf.~\eqref{symmetry+ both}], and the far-field condition \eqref{refined far}. However, matching shear stress \eqref{dist continuous stress outer} with the leading-order interior solution \eqref{dist w-1- solution} now provides an inhomogeneous Neumann condition
\begin{equation}
    \pd{w_0^+}{y} = \frac{\lambda_{-1}}{H} \quad \text{at} \quad y = 0.
\end{equation}
Since this shear stress contributes to the drag, the contribution from the corner singularity reduces from $1$ to $\lambda_{-1}$, so instead of \eqref{source} we have
\begin{equation}
    w_0^+ \sim \frac{2}{\pi}\lambda_{-1} \ln \rho \quad \text{as} \quad \rho \to 0.  
\end{equation}
Indeed, this condition can also be derived by matching via the intermediate region to the tip solution \eqref{omega+ lo}, whose amplitude is $\lambda_{-1}$.

Upon subtracting the shear $(\lambda_{-1}/H)y$ and the far-field constant $\lambda_0$ from $w_0^+$, we recover a rescaled form of the problem for $\tilde{w}_0^+$ that was solved in Appendix \ref{app:w0tilde+}. Hence, the required solution is
\begin{equation}\label{dist0 w0+ solution}
    w_0^+ = \lambda_{-1} \tilde{w}_0^+ + \frac{\lambda_{-1}}{H}y + \lambda_0,
\end{equation}
where $\tilde{w}_0^+$ is given by \eqref{w0tilde+ app result}. In particular, the local behavior near the tip is
\begin{equation}\label{dist0 w0+ near tip}
    w_0^+ \sim \lambda_{-1}\left(\frac{2}{\pi}\ln\rho + \beta\right) + \lambda_0,
\end{equation}
where $\beta$ is (the pure number) given by \eqref{beta}.

Next, on the intermediate scale at $\ord(1)$, the field $W_0^+(X,Y)$ satisfies Laplace's equation [cf.~\eqref{Laplace both}], homogeneous Neumann conditions at $X=0$ and $Y=0$ [cf.~\eqref{symmetry+ x=0} and \eqref{dist continuous stress intermediate}], and asymptotic matching to the outer solution \eqref{dist0 w0+ near tip} in the form
\begin{equation}
    W_0^+ \sim \lambda_{-1}\left[\frac{2}{\pi}\ln(\mu^{1/2}R) + \beta\right] + \lambda_0 \quad \text{as} \quad R \to \infty,
\end{equation}
where $R= \sqrt{X^2+Y^2}$. We conclude that the solution is simply an extrapolation of the outer solution \eqref{dist0 w0+ solution} [cf.~\eqref{uniform lo velocity intermediate}],
\begin{equation}\label{dist0 W0+ solution}
    W_0^+ = \lambda_{-1}\left[\frac{2}{\pi}\ln(\mu^{1/2}R) + \beta\right] + \lambda_0.
\end{equation}

Finally, on the tip scale at $\ord(1)$, the general solution \eqref{higher order tip C0} remains valid. Employing a 2--2 van Dyke matching between $\omega^+$ and $W^+$ [cf.~\eqref{2--2}], we determine the constant
\begin{equation}\label{dist0 Stip C0}
    C_0 = \lambda_{-1}\beta + \lambda_0.
\end{equation}
Hence, the result \eqref{higher order tip} for $\omega_0^\pm$ remains valid, except with a factor $\lambda_{-1}$ multiplying $\beta$.

\subsection{Correction to the outer solution in the interior domain}\label{app:dist0 Souter}

We now consider the $\ord(1)$ outer-scale problem in the interior domain. As  in the leading-order calculation [see \eqref{dist w-1- ode}--\eqref{dist w-1- bcs}], the field $w_0^-$ satisfies $\partial^2 w_0^-/\partial Y^2 = 0$ with the no-slip condition $w_0^- = 0$ at $Y=-H$. Here, the interfacial condition follows from continuity with \eqref{dist0 w0+ solution}, 
\begin{equation}\label{dist0 w0- bc interface}
    w_0^- = \lambda_{-1}\tilde{w}_0^+(x,0) + \lambda_0 \quad \text{at} \quad Y=0.
\end{equation}
The solution \eqref{dist w-1- solution} is then simply modified to
\begin{equation}\label{dist0 w0- solution}
    w_0^- = \left[\lambda_{-1} \tilde{w}_0^+(x,0) + \lambda_0\right]\left(1 + \frac{Y}{H}\right).
\end{equation}
Similarly to the leading-order calculation, the symmetry condition at $x=1$ is identically satisfied (given the symmetry of $\tilde{w}_0^+$), while the symmetry condition 
at $x=0$ does not need to be satisfied due to the presence of an intervening intermediate region.

\subsection{Leading-order intermediate-scale approximation in the interior domain}\label{app:dist0 intermediate}

We follow with the leading-order interior problem in the intermediate region. The field $W^-_{-1}$ satisfies Laplace's equation in the $(X,Y)$ coordinates [cf.~\eqref{Laplace both}] and the no-slip conditions [cf.~\eqref{noslip both}]
\begin{subequations}\label{dist0 W-1- bcs}
\refstepcounter{equation}
\refstepcounter{equation}
\begin{equation}\tag{\ref*{dist0 W-1- bcs}a,b}
    W^-_{-1}=0 \quad \text{at} \quad X=0, \qquad W_{-1}^- = 0 \quad \text{at} \quad Y = -H.
\end{equation}
Further, making use of the 
uniform exterior approximation \eqref{uniform lo velocity intermediate} and the velocity-continuity condition \eqref{continuous velocity} yields
\begin{equation}
    W^-_{-1}=\lambda_{-1} \quad \text{at} \quad Y=0,
\end{equation} 
\end{subequations}
while matching to the outer approximation \eqref{dist w-1- solution} requires
\begin{equation}\label{dist0 W-1- far}
    W_{-1}^- \sim \lambda_{-1}\left(1 + \frac{Y}{H}\right) \quad \text{as} \quad X \to \infty.
\end{equation}

As in \eqref{complex potential *} and \eqref{complex potential int}, we embed $W_{-1}^-$ in a complex potential,
\begin{equation}\label{complex potential intermediate}
	\Phi = W_{-1}^- + \ii \chi,
\end{equation}
which must be an analytic function of $Z= X + \ii Y$, in the horizontal semi-infinite strip $X > 0$, $-H < Y < 0$. We use the conformal map $\hat{z} = \ii (2Z/H) - 1$ to rotate and stretch this strip to the previously used semi-infinite strip (Fig.~\ref{fig:mapping}), which is then again mapped to the upper-half plane using $\tilde{z} = \sin(\pi \hat{Z}/2)$. Thus, the overall map is given by
\refstepcounter{equation}\label{dist0 intermediate coshmap}
\begin{equation}\tag{\ref*{dist0 intermediate coshmap}a,b,c}
	\tilde{z} = -\cosh \frac{\pi Z}{H}, \qquad \tilde{x} = -\cosh\frac{\pi X}{H} \cos \frac{\pi Y}{H}, \qquad \tilde{y} = -\sinh\frac{\pi X}{H}\sin\frac{\pi Y}{H}.
\end{equation}

The Dirichlet conditions \eqref{dist0 W-1- bcs} and the matching condition \eqref{dist0 W-1- far} become
\refstepcounter{equation}\label{W_-1^- conditions}
\begin{equation}
\tag{\ref*{W_-1^- conditions}a,b}
    W_{-1}^-(\tilde{x},0) = \left\{\begin{array}{lll} \lambda_{-1} & \text{ for } & \tilde{x} < -1, \\ 0 & \text{ for } & \tilde{x} > -1, \end{array}\right.
    \qquad
    W_{-1}^- \sim \lambda_{-1}\frac{\arg \tilde z}{\pi} \quad \text{as} \quad |\tilde{z}| \to \infty.
\end{equation}
The solution is thus given by
\refstepcounter{equation}\label{Phi W_-1^- solutions}
\begin{equation}
\tag{\ref*{Phi W_-1^- solutions}a,b}
    \Phi = \frac{\lambda_{-1}}{\ii \pi} \log(1+\tilde{z}), \qquad W_{-1}^- = \frac{\lambda_{-1}}{\pi} \arg(1+\tilde{z}).
\end{equation}
Reverting to $(X,Y)$ using \eqref{dist0 intermediate coshmap} gives
\begin{equation}
    W_{-1}^- = \frac{\lambda_{-1}}{\pi}\arctan \frac{\sinh(\pi X/H) \sin(\pi Y/H)}{\cosh (\pi X/H) \cos(\pi Y/H)-1},
\end{equation}
where the range of the arctangent is taken to be $0$ to $\pi$. The interfacial shear is then
\begin{equation}\label{dist0 intermediate shear}
    \left.\pd{W_{-1}^-}{Y}\right|_{Y=0} = \frac{\lambda_{-1}}{H}\coth \frac{\pi X}{2H}.
\end{equation}

\subsection{Drag and slip length}\label{app:dist0 balance}

We now calculate the contributions to the drag \eqref{balance split3} up to and including $\ord(\mu^{1/2})$. Using \eqref{dist0 Stip C0}, the tip contribution \eqref{balance split3 tip} is analogous to \eqref{Stip correction} but with factors of $\lambda_{-1}$ inserted:
\begin{equation}\label{dist0 Stip result}
    \Dtip \sim \lambda_{-1} + \mu^{1/2} \left(-\frac{2}{\pi}\lambda_1 \ln\frac{1}{\delta_1} + \lambda_{-1}\beta + \lambda_0\right).
\end{equation}
The intermediate contribution \eqref{balance split3 intermediate} is, using \eqref{dist0 intermediate shear}, 
\begin{subequations}
\begin{align}
    \Dintermediate &\sim \mu^{1/2} \int_{\delta_1/\mu^{1/2}}^{\delta_2/\mu^{1/2}} \frac{\lambda_{-1}}{H}\coth \frac{\pi X}{2H}\,\intd{X}
    \\
    &= \mu^{1/2} \frac{2\lambda_{-1}}{\pi}\ln \frac{\sinh[\pi \delta_2/(2H\mu^{1/2})]}{\sinh[\pi \delta_1/(2H\mu^{1/2})]}.
\end{align}
\end{subequations}
Since $\delta_1 \ll \mu^{1/2} \ll \delta_2$ [see \eqref{delta3 range}], we can expand this as
\begin{equation}\label{dist0 Sintermediate result}
    \Dintermediate \sim \lambda_{-1}\left(\frac{\delta_2}{H} + \mu^{1/2}\frac{2}{\pi}\ln \frac{H \mu^{1/2}}{\pi \delta_1}\right).
\end{equation}
Finally, for the outer contribution \eqref{balance split3 outer}, the solutions \eqref{dist w-1- solution} and \eqref{dist0 w0- solution} yield
\begin{equation}
    \Douter \sim \int_{\delta_2}^{1} \left[\frac{\lambda_{-1}}{H} + \mu^{1/2} \frac{\lambda_{-1}\tilde{w}_0^+(x,0) + \lambda_0}{H}\right]\,\intd{x}.
\end{equation}
Using \eqref{w0tilde+ app result}, we find that $\int_0^1 \tilde{w}_0^+(x,0)\,\intd{x} = 0$, so we are left with
\begin{equation}\label{dist0 Souter result}
    \Douter \sim \frac{\lambda_{-1}}{H}(1 -\delta_2) + \mu^{1/2} \frac{\lambda_0}{H}.
\end{equation}

Adding together \eqref{dist0 Stip result}, \eqref{dist0 Sintermediate result} and \eqref{dist0 Souter result}, the terms involving $\delta_1$ and $\delta_2$ cancel as expected, and we obtain the total drag
\begin{equation}
    \Dtotal \sim \lambda_{-1}\left(1 + \frac{1}{H}\right) + \mu^{1/2} \left[\lambda_{-1} \frac{2}{\pi}\ln(H\mu^{1/2}) + \lambda_0\left(1 + \frac{1}{H}\right)\right],
\end{equation}
where the error is algebraically small. 
Requiring that the $\ord(\mu^{1/2})$ term vanishes [cf.~\eqref{balance split3 total}], we conclude that
\begin{equation}\label{lam0 dist 2-term app}
    \lambda_0 = \frac{2\lambda_{-1}}{\pi} \left(\frac{H}{1+H}\right) \ln \frac{1}{\mu^{1/2}H}.
\end{equation}
Substitution of \eqref{lam-1 dist} into \eqref{lam0 dist 2-term app} furnishes the requisite formula \eqref{lam dist 2-term}.

\bibliography{refs}
\end{document}